\documentclass[11pt,a4paper]{article}
\usepackage{fullpage}
\usepackage{lscape}
\usepackage{amssymb,amsmath,stmaryrd,amsthm}
\usepackage[mathscr]{eucal}
\usepackage{tikz} 
\usetikzlibrary{arrows,shapes,positioning,shadows,trees,chains,plotmarks}
\usetikzlibrary{decorations.pathreplacing}
\usepackage{color}
\usepackage{colortbl}

\theoremstyle{definition}
\newtheorem{definition}{Definition}

\newtheorem{proposition}{Proposition}
\newtheorem{lemma}{Lemma}
\newtheorem{corollary}{Corollary}

\newtheorem{observation}{Observation}

\theoremstyle{remark}

\def \NN {\mathbb{N}}
\def \PP {\mathbb{P}}

\def \cC {\mathcal{C}}

\def \cO {\mathcal{O}}

\def \1{{\mathbf{1}}}

\def \Max{\mathsf{Max}}
\def \Min{\mathsf{Min}}
\def \Pr{\mathrm{Pr}}

\def \prec{\mathsf{prec}}
\def \succ{\mathsf{succ}}
\def \dd{\mathrm{d}}
\def \bmu{\boldsymbol{\mu}}
\def \OS{\mathrm{OS}}
\def \Beta{\mathrm{Beta}}

\begin{document}
\title{An approximation algorithm for random generation of capacities}
\author{Michel GRABISCH${}^{1}$, Christophe
  LABREUCHE${}^2$ and Peiqi SUN${}^3$\thanks{Corresponding author.}\\
\normalsize $^{1}$ Paris School of Economics, Universit\'e Paris I - Panth\'eon-Sorbonne, Paris, France  \\
{\normalsize\tt michel.grabisch@univ-paris1.fr}\\
\normalsize $^{2}$ Thales Research \& Technology, Palaiseau, France\\
{\normalsize\tt christophe.labreuche@thalesgroup.com}\\
\normalsize $^{3}$ Universit\'e Paris I - Panth\'eon-Sorbonne, Paris, France  \\
{\normalsize\tt peiqisun94@gmail.com}}

\date{\today}
\maketitle

\begin{abstract}
  Capacities on a finite set are sets functions vanishing on the empty set and
  being monotonic w.r.t. inclusion. Since the set of capacities is an order
  polytope, the problem of randomly generating capacities amounts to generating
  all linear extensions of the Boolean lattice. This problem is known to be
  intractable even as soon as $n>5$, therefore approximate methods have been
  proposed, most notably one based on Markov chains. Although quite accurate,
  this method is time consuming. In this paper, we propose the 2-layer
  approximation method, which generates a subset of linear extensions,
  eliminating those with very low probability. We show that our method has
  similar performance compared to the Markov chain but is much less time
  consuming. 
\end{abstract}


\section{Introduction}
Capacities, introduced by Choquet \cite{cho53}, are monotone set functions
vanishing on the empty set. They are widely used in decision making under risk
and uncertainty, multicriteria decision making, and related to many
combinatorial problems in Operations Research (see \cite{gra16} for a detailed
account).

In the phase of determining or learning a model based on capacities, it is often
important to generate randomly capacities in a uniform way. The problem appears
to be surprisingly difficult, due to the monotonicity constraints, and naive
methods give poor results, with extremely biased distributions for some
coefficients of the capacity. The theoretical exact solution to this problem is
however known, since the set of capacities on a finite set $N$ is an order
polytope, associated to the Boolean lattice $(2^N,\subseteq)$, and generating
uniformly random elements in an order polytope amounts to generating all linear
extensions of the the underlying poset, that is, in our case, the Boolean
lattice. Unfortunately, the number of linear extensions in $(2^N,\subseteq)$
grows tremendously with $|N|$, and is even not known beyond $|N|=7$. This
obliges to resort to approximate methods. The literature has produced many such
methods, among which the Markov chain approach seems to give the best results.

The aim of this paper is to propose a new approximate method, counting a limited
number of linear extensions, eliminating linear extensions with very low
probability. The basic idea for this is to concentrate on the two top and two
bottom layers on the Boolean lattice, and to eliminate one by one maximal
elements in the two top layers, and minimal elements in the two bottom
layers. For this reason, the method is called {\it 2-layer approximation
  method}.

A second contribution of the paper is to provide two ways for measuring
the performance of any method generating uniformly distributed random
capacities. The first  way is based on the probability distribution of the
coefficients of the capacity. It is shown that if a capacity $\mu$ is uniformly
generated, then the distribution of $\mu(S)$ for any $S\subseteq N$ depends only
on the cardinality of $S$, and is symmetric w.r.t. the distribution of
$\mu(N\setminus S)$. However, the exact form of the distribution of $\mu(S)$
seems to be very difficult to compute.  A second way consists in computing the
centroid of the set of generated capacities, and to verify that it is close to
the theoretical centroid of the polytope of capacities. Our experiments show
that our 2-layer approximation method slightly outperforms the Markov chain
method, while taking much less computation time.

The paper is organized as follows. Section~\ref{sec:prel} settles the problem in
its full generality, relating uniform random generation with linear
extensions. Section~\ref{sec:rangen} focuses on the generation of capacities and
gives the state of the art. The 2-layer approximation method is developed in
Section~\ref{sec:2l} in full detail. Section~\ref{sec:perf} is devoted to the
measure of performance of any method of random generation of capacities.

\section{Preliminaries}\label{sec:prel}
Let $P$ be a finite set, endowed with a partial order $\preccurlyeq$, that is,
a reflexive, antisymmetric and transitive binary relation. We say that
$(P,\preccurlyeq)$ is a {\it  (finite) poset}.
Recall that $x\in P$ is {\it
  maximal} if $x\preccurlyeq y$ with $y\in P$ implies $x=y$. Similarly, $x\in P$
is {\it minimal} if $y\preccurlyeq x$ with $y\in P$ implies $x=y$. We denote by
$\Max(P,\preccurlyeq)$ and $\Min(P,\preccurlyeq)$ (simply $\Max(P), \Min(P)$ if
there is no ambiguity) the set of maximal and minimal
elements of $P$, respectively.


The {\it order polytope} \cite{sta86}
associated to $(P,\preccurlyeq)$, denoted by $\cO(P)$, is the set
\[
\cO(P) = \{f:P\longrightarrow [0,1]\mid f(x)\leqslant f(y)\text{ if }
x\preccurlyeq y\}.
\]
It is a polytope of dimension $p:=|P|$, whose volume $V(\cO(P))$ can be computed
by the following formula \cite{sta86}:
\begin{equation}\label{eq:1}
V(\cO(P)) = \frac{e(P)}{p!},
\end{equation}
where $e(P)$ is the number of linear extensions of $(P,\preccurlyeq)$. A {\it
  linear extension} of $(P,\preccurlyeq)$ is a total order $\leqslant$ on $P$ which
is compatible with the partial order $\preccurlyeq$ in the following sense:
$x\preccurlyeq y$ implies $x\leqslant y$. It is convenient to denote a linear
extension by the sequence of the elements of $P$ arranged in increasing order
according to $\leqslant$: assuming $P=\{x_1,\ldots,x_p\}$, the linear extension
$\leqslant$ is denoted by $x_{\sigma(1)},\ldots,x_{\sigma(p)}$, with $\sigma$ a
permutation on $\{1,\ldots,p\}$ and $x_{\sigma(1)}<\cdots< x_{\sigma(p)}$.

\medskip

Let $N:=\{1,2,\ldots,n\}$ be a finite set of $n$ elements. A {\it (normalized)
  capacity} \cite{cho53,sug74,gra16} on $N$ is a set function
$\mu:2^N\longrightarrow [0, 1]$ satisfying $\mu(\varnothing)=0$, $\mu(N)=1$
(normalization), and the property $S\subseteq T\Rightarrow \mu(S)\leqslant
\mu(T)$ (monotonicity).

We denote the set of capacities on $N$ by $\cC(N)$. From its definition, one can
see that $\cC(N)$ is an order polytope, whose underlying poset is
$(2^N\setminus\{\varnothing,N\},\subseteq)$.

\section{Random generation of capacities}\label{sec:rangen}
\subsection{Random generation and linear extensions}\label{sec:rgline}
The problem of randomly generating capacities according to a uniform
distribution amounts to picking a point in the polytope $\cC(N)$ in a uniform
way. This is made simple because $\cC(N)$ is an order polytope. According to
Stanley \cite{sta86}, given a poset $(P,\preccurlyeq)$ with
$P=\{x_1,\ldots,x_p\}$ and its associated order polytope $\cO(P)$, each linear
extension of $(P,\preccurlyeq)$ defines a region in $\cO(P)$:
\[
R_\sigma:=\{f\in\cO(P)\mid 0\leqslant f(x_{\sigma(1)})\leqslant
f(x_{\sigma(2)})\leqslant\cdots\leqslant f(x_{\sigma(p)})\leqslant 1\}
\]
where $\sigma$ is the permutation associated to the linear extension, i.e.,
$x_{\sigma(1)}<\cdots< x_{\sigma(p)}$. All regions $R_\sigma$
are identical (up to a change of coordinates), are $p$-dimensional simplices
with volume $\frac{1}{p!}$, which leads to (\ref{eq:1}).
Vertices of this simplex are the $p+1$ functions given by
\begin{equation}\label{eq:1a}
0=f(x_{\sigma(1)})=\cdots= f(x_{\sigma(k)}),
\ f(x_{\sigma(k+1)})=\cdots=f(x_{\sigma(p)})= 1, \quad (k=0,1,\ldots,p).
\end{equation}

As a consequence, the
random generation of a point in $\cO(P)$ w.r.t. a uniform distribution amounts
to uniformly selecting a linear extension, and then to uniformly selecting a point in
the associated region $R_\sigma$. The latter step is done as follows: generate
$p$ independent numbers in $[0,1]$ according to the uniform distribution, then
order them in increasing order, say, $z_1\leqslant\cdots\leqslant z_p$, and put
$f(x_{\sigma(1)})=z_1,\ldots,f(x_{\sigma(p)})=z_p$. This define $f\in\cO(P)$.

\medskip

The above technique can be applied to the uniform generation of capacities. As
explained above, $(2^N\setminus\{\varnothing,N\},\subseteq)$ is the underlying
poset. It suffices to select randomly a linear extension $<$ of
$(2^N\setminus\{\varnothing,N\},\subseteq)$, say
\[
S_{\sigma(1)}<\cdots <S_{\sigma(2^n-2)},
\]
and then generate independently and uniformly $2^n-2$ numbers in
$[0,1]$, letting $\mu(S_{\sigma(i)})=z_i$, where $z_i$ is the $i$th smallest
generated number. The problem is then theoretically solved but the above method
reveals to be intractable as soon as $n\geqslant 5$. Indeed, the number of
linear extensions on the Boolean lattice $(2^N,\subseteq)$ (note that it is equal
to the number of linear extensions on
$(2^N\setminus\{\varnothing,N\},\subseteq)$) increases extremely fast, as
indicated in Table~\ref{tab:1} below.
\begin{table}[htb]
  \begin{center}
    \begin{tabular}{|c|r|}\hline
    $n$ & $e(2^N)$ \\ \hline
    1 & 1\\
    2 & 2\\
    3 & 48\\
    4 & 1680384\\
    5 & 14807804035657359360\\
    6 & 141377911697227887117195970316200795630205476957716480\\ \hline
    \end{tabular}
    \caption{Number of linear extensions of $2^N$}
  \label{tab:1}
  \end{center}
\end{table}
There is no known closed-form formula for computing the number of linear
extensions on $2^N$, nor is it known beyond $n=7$ (it is sequence A046873 in the
{\it Online Encyclopedia of Integer Sequences}), however bounds are known (see,
e.g., \cite{brte03}). More generally, the problem of counting the number of
linear extensions on a poset is a $\sharp$-P complete problem, as shown in
\cite{brwi91}.

As a consequence, only approximate methods can be used to generate capacities in
a uniform way.

\subsection{Related literature}
We give here a brief account of the literature on approximate methods for
generating capacities. One of the simplest method (shown in \cite{hapi17}) is to
select randomly in each step one element of a capacity and then draw uniformly a
number within an interval imposed by the monotonicity constraints of the already
drawn numbers. This method, called {\it raandom node generator}, although very
simple to implement, yields fairly biased results.

Several methods try to generate linear extensions of the considered poset, from
which it is immediate to generate a capacity as we explained above. Karzanov and
Khachiyan \cite{kakh91} propose to use the Markov Chain technique to generate
linear extensions of a given poset $P$. Consider each linear extension as a
state of a Markov Chain, and denote by $E$ the set of all linear extensions of
$P$. The Markov chain with its transition probability matrix describes a random walk
through the simplices of the order polytope $\mathcal{O}(P)$, and Karzanov and
Khachiyan prove that its transition probability is an ergodic time-reversible
Markov chain with a uniform stationary distribution, which means that for an
arbitrary initial probability distribution on $E$, after $T$ steps ($T$ large
enough), the distribution converges to the uniform distribution on $E$. Thus,
for any given poset $P$ and sufficiently large $T$, the Markov chain method
gives a nearly uniform generator of linear extensions of the poset. However,
when the cardinality of $P$ is large, it is costly in computation time. In
addition to the original classical Karzanov-Khachiyan chain method, there exist
some more efficient Markov Chain algorithms, for example, Bubley and Dyer
introduced the Insertion Chain algorithm in \cite{budy91}, while \cite{taniko17}
mentions some other Markov Chain algorithms.

Combarro et al. \cite{codimi13} gives a good survey of methods of generating
capacities, and propose a method to generate linear extensions of $P$, which
consists in selecting a minimal element of $P$ with a certain
probability, then delete it and repeat this procedure until all elements have been
selected. If we could determine this probability exactly, we would obtain an
exact method. For example, for generating 2-symmetric capacities, Miranda and
Garcia Segador \cite{miga20} transform the problem into generating a
linear extension of the Young rectangular diagram. Since there exists a
closed-form formula for the number of linear extensions of the Young diagram,
according to the general probability formula of an element $x$ being selected $
\mathbb{P}(x) = \frac{e(P\setminus x)}{e(P)}$ (see Formula (\ref{eq:2}) below), we get an exact method to
generate 2-symmetric capacities. Besides, Beliakov introduces an
approximation method in \cite{bel22} to generate supermodular capacities, and
Miranda and Garcia Segador introduce an exact method in \cite{miga19} to
generate 2-additive capacities.

For a general capacity, the idea of generating a linear extension by the method
introduced in \cite{codimi13} could be used, at the condition to find a way to
determine the probability of selecting an element. Miranda and Garcia Segador
have proposed some approximation methods to obtain this probability, namely the
Bottom-up method \cite{miga19a}, while Combarro et al. have proposed the
Minimals Plus method \cite{cohudi19}.

In this article, we present another approximation method to determine the
probability of selecting a minimal (or maximal) element, which is presented in
the next section.

\section{The 2-layer approximation algorithm}\label{sec:2l}
\subsection{Basic idea}\label{sec:bi}

Let us consider an arbitrary poset $(P,\preccurlyeq)$ with
$P=\{x_1,\ldots,x_p\}$. Clearly, any linear extension
$x_{\sigma(1)},\ldots,x_{\sigma(p)}$ starts with a minimal element and terminates
with a maximal element, i.e., $x_{\sigma(1)}\in\Min(P)$ and $x_{\sigma(p)}\in\Max(P)$. 
Observe that $x_{\sigma(2)}$ must be a minimal element of
$P\setminus\{x_{\sigma(1)}\}$, and $x_{\sigma(p-1)}$ is a maximal element on
$P\setminus\{x_{\sigma(p)}\}$, and so on. This gives a recursive structure to
linear extensions.

Based on this observation, the probability that a linear extension of $P$ starts
with $m\in\Min(P)$ is
\begin{align}
  \Pr(m\mid P) &= \frac{e(P\setminus\{m\})}{e(P)}\nonumber \\
          &= 1-\frac{\sum_{m'\in\Min(P), m'\neq m}e(P\setminus\{m'\})}{e(P)},\label{eq:2}
\end{align}
and smilarly the probability that a linear extension terminates with
$M\in\Max(P)$ is
\begin{align}
  \Pr(M\mid P) &= \frac{e(P\setminus\{M\})}{e(P)}\nonumber\\
          &= 1-\frac{\sum_{M'\in\Max(P), M'\neq M}e(P\setminus\{M'\})}{e(P)}.\label{eq:3}
\end{align}
Then, generating a linear extension amounts to choosing, according to the
correct probability given by (\ref{eq:2}) and (\ref{eq:3}), a minimal or a
maximal element of a poset which is diminished by one element at each step. As
the computation of the probability directly depends on the number of linear
extensions, the computation can be exact only when $P$ becomes small
enough. Otherwise, some approximation must be done. The idea we propose is to
take the lower part of the poset for choosing minimal elements, and the upper part for
choosing maximal elements, thus neglecting minimal and maximal elements which
are outside these two subparts. These two subposets are also used to make the
computation of the respective probabilities by (\ref{eq:2}) and
(\ref{eq:3}). The method, explained in the next subsection, is however specific
to posets being subsets of the Boolean lattice $(2^N,\subseteq)$.

\subsection{Approximation method}
From now on, we deal with posets which are particular subsets of
$(2^N\setminus\{\varnothing,N\},\subseteq)$. As explained in Section~\ref{sec:bi}, we proceed by deleting
step by step minimal or maximal elements of the Boolean lattice
$(2^N\setminus\{\varnothing,N\},\subseteq)$, which yields at each step of the process a poset $(H,\subseteq)$ which is
a subset of $(2^N\setminus\{\varnothing,N\},\subseteq)$. For commodity, and referring to its Hasse
diagram, we call {\it nodes} the subsets of $N$ in $H$, denoting them most
often by $x,y,\ldots$ (unless the usual set notation $S,T,\ldots$ is more
adequate), and we define a {\it layer} of $H$ as the set of all nodes (subsets)
of same cardinality.

\begin{figure}[htb]
\centering\small
\begin{tikzpicture}[scale = 1.5]
\node[right] at (-1.1,1-0.05){$\{1\}$};\node[right] at (0-0.1,1-0.05){$\{2\}$};
\node[right] at (1-0.1,1-0.05){$\{3\}$};\node[right] at (2-0.1,1-0.05){$\{4\}$};\node[right] at (0.5-0.1,0-0.05){$\emptyset$};\draw[-=0.5](0.5,0) -- (0,1-0.25); \draw[fill] (0.5,0) circle (0.04);
\draw[-=0.5](0.5,0) --(-1,1-0.25);\draw[-=0.5](0.5,0) -- (1,1-0.25);
\draw[-=0.5](0.5,0) -- (2,1-0.25);\draw[-=0.5](0,1-0.25) -- (-2,2-0.25);

\node[right] at (-2-0.1,2-0.05){$\{12\}$};\node[right] at (-1-0.1,2-0.05){$\{13\}$};
\node[right] at (0-0.1,2-0.05){$\{14\}$};\node[right] at (1-0.3,2-0.05){$\{23\}$};
\node[right] at (3-0.3,2-0.05){$\{34\}$};\node[right] at (2-0.1,2-0.05){$\{24\}$};
\node[right] at (0.1,3-0.25){$\{124\}$};\node[right] at (2.1,3-0.25){$\{234\}$};
\draw[-=0.5](0,1-0.25) -- (-2,2-0.25);\draw[fill] (0.0,0.75) circle (0.04);();
\draw[-=0.5](0,1-0.25) -- (2,2-0.25);\draw[-=0.5](0,1-0.25) -- (1,2-0.25);
\draw[-=0.5](-1,1-0.25) -- (-2,2-0.25);\draw[-=0.5](-1,1-0.25) -- (-1,2-0.25);
\draw[-=0.5](-1,1-0.25) -- (0,2-0.25);\draw[fill] (-1,0.75) circle (0.04);();
\draw[-=0.5](1,1-0.25) -- (-1,2-0.25);\draw[-=0.5](1,1-0.25) -- (1,2-0.25);
\draw[-=0.5](1,1-0.25) -- (3,2-0.25);\draw[fill] (1,0.75) circle (0.04);();
\draw[-=0.5](2,1-0.25) -- (3,2-0.25);\draw[-=0.5](2,1-0.25) -- (2,2-0.25);\draw[-=0.5](2,1-0.25) -- (0,2-0.25);
\draw[fill] (2,0.75) circle (0.04);();\draw[color = red,fill] (-1,1.75) circle (0.04);();
\draw[fill] (0,1.75) circle (0.04);();\draw[fill] (-2,1.75) circle (0.04);();
\draw[fill] (1,1.75) circle (0.04);();\draw[fill] (2,1.75) circle (0.04);();
\draw[fill] (3,1.75) circle (0.04);();\draw[color = red,fill] (0,2.75) circle (0.04);();
\draw[color = red,fill] (2,2.75) circle (0.04);();\draw [very thick, color = red, dotted] (-2.2,1.6) rectangle (3.3,3);
\draw[-=0.5](-2,2-0.25) -- (0,3-0.25);\draw[-=0.5](2,2-0.25) -- (0,3-0.25);
\draw[-=0.5](2,2-0.25) -- (2,3-0.25);\draw[-=0.5](1,2-0.25) -- (2,3-0.25);
\draw[-=0.5](3,2-0.25) -- (2,3-0.25);\draw[-=0.5](0,2-0.25) -- (0,3-0.25);
\end{tikzpicture} 
\hfill
\caption{Example of poset $H$ with the poset $T_H(2,6,1)$ formed by the two top
  layers framed in red. Braces and commas are omitted for denoting sets. Maximal
  elements of $H$ are in red. 13 is the only isolated node.}
\label{fig:1}
\end{figure}
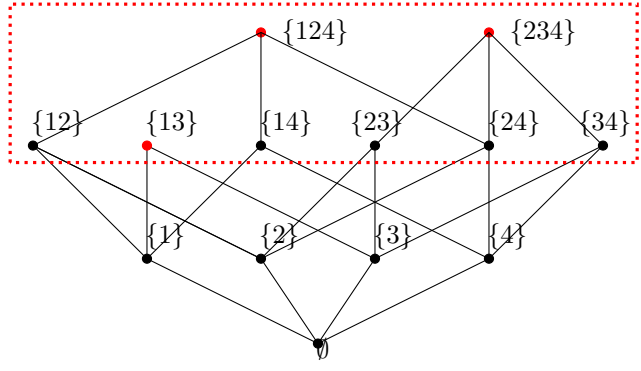

The principle of the method is to limit the search of maximal elements (resp.,
minimal elements) of $H$ as well as the computation of their probability of
occurrence to the two top layers (resp., to the two bottom layers).  We denote
by $T_H$ the poset formed by the two top layers of $H$, and for ease of
readability often write $T_H[h,k,|I|]$ to specify its {\it basic parameters}
$h,k,|I|$ explained below.  The upper layer contains $h$ nodes of cardinality
say $\ell$, with $2\leqslant \ell\leqslant n-1$, while the lower layer contains
$k$ nodes of cardinality $\ell-1$, among which some of them are isolated, i.e.,
with no predecessor, and we denote by $I$ the set of isolated nodes and by $|I|$
its number. A node $y$ of the lower layer has {\it predecessors} in the upper
layer, i.e., sets $x$ satisfying $x\supset y$. We denote by $\prec(y)$ the set
of predecessors of $y$. Similarly, for a node $x$ of the upper layer, we denote
by $\succ(x)$ the set of its {\it successors}, i.e., subsets of $x$ in the lower
layer.  In a dual way, we introduce $B_H$ (denoted also $B_H[h',k',|I'|]$), the
poset of the two bottom layers of $H$, with $h'$ nodes on the upper level, $k'$
nodes on the lower level, and $I'$ is the set of isolated nodes in the upper (!)
level.  Fig.~\ref{fig:1} illustrates the definitions.

Consider $H\subseteq 2^N\setminus\{\varnothing,N\}$ and take a maximal element $M$ of $H$ belonging to
$T_H$ (i.e., a maximal element of $T_H$), and a minimal element $m$ of $H$
belonging to $B_H$ (note that it is equivalent to write $T_H\setminus\{M\}$ or
$T_{H\setminus\{M\}}$).  The probabilities that $M$ and $m$ are terminal and
initial elements of a linear extension of $H$ are approximated by:
\begin{align}
  \Pr(M\mid H) \approx &
  \frac{e(T_H\setminus\{M\})}{e(T_H)}=\Pr(M\mid T_H) \label{eq:4}\\
  \Pr(m\mid H) \approx &
  \frac{e(B_H\setminus\{m\})}{e(B_H)}=\Pr(m\mid B_H).\label{eq:5} 
\end{align}
This approximation is based on the following observation: the number of nodes
which are above a given node in layer $k$ increases very fast with $k$. Indeed,
supposing that a node has in average $\ell$ predecessors, each predecessor has
in turn $\ell$ predecessors as well, so that the number of nodes above a given one
in the two layers above it is of the order $\ell^2$, and so on. Consequently, a
node $x$ in the third layer (or below)  has very little probability to become a
maximal element and hence to be selected,
since all nodes above it must be eliminated first, {\it without eliminating all
  nodes of the 1st layer}, because otherwise $x$ becomes a node of the 2nd
layer.

The algorithm for the generation of linear extensions with uniform probability is
described below.
\begin{quote}
  {\bf generate-linext}$(P,l)$\\
  {\bf Input:} a poset $P$ subset of $2^N\setminus\{\varnothing,N\}$\\
  {\bf Output:} a linear extension $l$ of $P$ generated with a uniform
  distribution\\
  $H\leftarrow P$; $lmin\leftarrow \varnothing$; $lmax\leftarrow\varnothing$\\
  {\bf While} height of $H> 2$ {\bf do}\\
  \hspace*{1cm} Compute the basic parameters of $T_H$: $k,h,|I|$\\
  \hspace*{1cm} Select $M\in T_H[h,k,|I|]$ with probability   $\Pr(M\mid  T_H[h,k,|I|])$\\
  \hspace*{1cm} Add $M$ at the beginning of $lmax$\\
  \hspace*{1cm} Compute the basic parameters of $B_H$: $k',h',|I'|$\\
  \hspace*{1cm} Select $m\in B_H[h',k',|I'|]$ with probability   $\Pr(m\mid B_H[h',k',|I'|])$\\
  \hspace*{1cm} Add $m$ at the end of $lmin$\\
  $H\leftarrow H\setminus\{M,m\}$\\
          {\bf end while}\\
          \% Now $H$ is reduced to two layers: $B_H$ and $T_H$ coincide\\
  {\bf While} height of $H= 2$ {\bf do}\\
  \hspace*{1cm}{\bf If} number of nodes in the upper layer $\leqslant$ number of nodes in the
  lower layer {\bf then}\\
  \hspace*{2cm}  Select $M\in T_H[h,k,|I|]$ with probability   $\Pr(M\mid
  T_H[h,k,|I|])$\\
  \hspace*{2cm} Add $M$ at the beginning of $lmax$\\
  \hspace*{1cm}{\bf otherwise}\\
  \hspace*{2cm}  Select $m\in B_H[h',k',|I'|]$ with probability   $\Pr(m\mid
  B_H[h',k',|I'|])$\\
  \hspace*{2cm} Add $m$ at the end of $lmin$\\
  \hspace*{1cm} {\bf end if}\\
  {\bf end while}\\
  \% Now $H$ is reduced to one layer, which is an antichain whose elements
  have\\ 
  \% the same probability\\
  {\bf While} $H\neq\varnothing$ {\bf do}\\
     \hspace*{1cm} Select uniformly at random an element $x\in H$\\
     \hspace*{1cm} Add $x$ at the end of $lmin$\\
     \hspace*{1cm} $H\leftarrow H\setminus\{x\}$\\
  {\bf end while}\\
  $l\leftarrow lmin$ ; concatenate $lmax$ to the end of $l$           
\end{quote}
Observe the particularity when the height of $H$ is equal to 2. Indeed, the two
top layers coincide with the two bottom layers, so that one may either select
maximal or minimal elements. In order to minimize the number of steps, $H$ is
considered as $B_H$ if there are more nodes in the upper layer, and as $T_H$
otherwise.

\medskip

We now aim at getting an explicit expression of $\Pr(M\mid T_H[h,k,|I|])$ (the same development can be done in a dual way for $\Pr(\mid B_H[h',k',|I'|])$). 

We start with a simple property and observation.
\begin{lemma}\label{lem:1}
  For any poset $H\subseteq 2^N\setminus\{\varnothing,N\}$, the number of linear extensions in the poset
  $T_H[h,k,|I|]$ is
  \[
  e(T_H(h,k,|I|)) = e(T_{H\setminus I}[h,k-|I|,0]) \times \prod_{i=1}^{|I|}(k-|I|+h+i).
 \]
\end{lemma}
\begin{proof}
  Take $H\subseteq 2^N\setminus\{\varnothing,N\}$ and consider the two top layers $T_H[h,k,|I|]$. The
  isolated nodes being incomparable with the remaining elements, they can appear
  at any position in a linear extension of $T_H[h,k,|I|]$. Hence, there are
  $k+h-|I|+1$ possibilities to insert one isolated node into a linear extension
  of $T_{H\setminus I}[h,k-|I|,0]$, then $k+h-|I|+2$ possibilities to
  insert a second one independently of the first choice, etc. This yields the
  desired formula.
\end{proof}

\medskip

Deriving other properties of $e(T_H)$ requires some assumption on the structure
of $T_H$. The following definition is central in our development.
\begin{definition}\label{def:nx}
  Let $x$ be a node of the upper layer of $T_H$.
  \begin{enumerate}
  \item The function $f_x$ assigns to every
node $y$ of the lower layer an integer as follows:
\[
f_x(y) = \begin{cases} |\prec(y)|, & y\in\succ(x)\\
  0, & \text{otherwise.}
  \end{cases}
\]
\item The function $n_x:\NN\rightarrow \NN$ is defined from $f_x$ as follows:
  $n_x(r)$ is the number of occurences of $f_x(y)=r$, i.e.,
  $n_x(r)=|f_x^{-1}(r)|$. When $r>0$, it is the number of successors of $x$
  having $r$ predecessors, otherwise when $r=0$ it is the number of nodes in the
  lower layer which are not successors of $x$.
  \end{enumerate}
\end{definition}
Definitions for $B_H$ are dual: $x$ is in the lower layer and for every $y$ in
the upper layer $f_x(y)$ is the number of successors of $y$, etc. 

\begin{definition}
We say that $T_H$ (respectively, $B_H$) is {\it regular} if $n_x$ is invariant with $x$, i.e., $n_x(r)=n_{x'}(r)$ for every $r$ and every two nodes $x,x'$ in the
upper layer (respectively, in the lower layer).
\end{definition}

As regularity will be the key property in the sequel, we elaborate on it and
relate it to more intuitive properties. The first one is based on the following
observation.
\begin{observation}
For any two nodes $x,x'$ in the upper level of poset $T_H$, we always have
$|\succ(x)\cap\succ(x')|\leqslant 1$. 
\end{observation}
\begin{proof}
let $x$ be in the upper layer of $T_H$ and suppose that $x$ has at least two
successors. As successors of $x$ have one less element than $x$, two successors
of $x$ have the form $S\cup\{i\}$ and $S\cup\{j\}$, with
$x=S\cup\{i,j\}$. Consequently, $S\cup\{i\}$ and $S\cup\{j\}$ cannot be successors of
another node $x'$ of the supper layer of $T_H$, hence $|\succ(x)\cap\succ(x')|\leqslant 1$. 
\end{proof}

We say that $T_H$ is {\it closed under intersection} if for every two
  nodes $x,x'$ of the upper layer, $|\succ(x) \cap \succ(x')|= 1$. i.e., $x\cap
  x'$ belongs to the lower layer.
Similarly for the two bottom layers, $B_H$ is {\it closed under union} if for
every node $x,x'$ of the lower layer, $x\cup x'$ belongs to the upper
layer. See the Appendix for a characterization of posets closed under
intersection. 

In addition, we say that $T_H$ is {\it balanced} if $|\succ(x)|=|\succ(x')|$ for
every $x,x'$ in the upper layer. Similarly, $B_H$ is balanced if
$|\prec(x)|=|\prec(x')|$ for every $x,x'$ in the lower layer.

The next proposition relates all these properties.
\begin{proposition}\label{prop:reg}
Consider a poset $H$.
\begin{enumerate}
\item If $T_H$ is regular, then it is balanced.
\item If $T_H$ is closed under intersection and balanced, then it is regular.
\item Regularity of $T_H$ does not imply that $T_H$ is closed under
  intersection.
\item $T_H$ is regular if $H=2^N\setminus\{N,\varnothing\}$.
\end{enumerate}
The same assertions hold for $B_H$.
\end{proposition}
(see proof in the Appendix)

Thus, in the algorithm {\bf generate-linext}, the alternance
of selecting maximal elements and minimal elements permits to keep as long as
possible the assumption of regularity.

\medskip

We are now in
position to show the next important result.
\begin{proposition}\label{prop:1}
Consider a poset $H$ with its two top layers $T_H[h,k,0]$  without
  isolated nodes, and suppose that it is regular. Then, for
  every $x$ in the upper layer,
  \begin{equation}\label{eq:6}
e(T_H[h,k,0]) = h\times e(T_{H\setminus\{x\}}[h-1,k,|I|]),
\end{equation}
where $I$ is the set of isolated nodes in the two upper layers of $H\setminus\{x\}$.
\end{proposition}
\begin{proof}
If $n_x$
is invariant w.r.t $x$, it means that every node $x$ has the same structure of
successors, implying that $H\setminus\{x\}$ has a number $I$ of isolated nodes
which does not depend of $x$, which in turn implies that
$e(T_{H\setminus\{x\}}[h-1,k,|I|])$ does not depend on $x$. As any node of the upper
layer is a maximal element, i.e., can be the last element in a linear extension,
this proves (\ref{eq:6}).
\end{proof}

The next step is to establish the probability to select a maximal element from
$T_H$. A maximal element can be either a node from the upper layer, or
an isolated node of the lower layer. It is easy to see that in the latter case,
the probability does not depend on the isolated element. This is because an
isolated node can appear in any position of the linear extension, without
constraint. Interestingly, one can prove that under regularity, the same
property holds for nodes of the upper layer.
\begin{proposition}\label{prop:2}
Consider a poset $H$ with its two top layers $T_H$, and suppose that it
is regular. Then, for
every node $x$ of the upper layer in $T_H$, the
probability $\Pr(x\mid T_H)$ that a
linear extension in $T_H$ terminates by $x$ is
independent of $x$.
\end{proposition}
\begin{proof}
We just need to prove that the poset $T_H$ is symmetric w.r.t. the nodes in the
upper layer, that is, all nodes in the upper layer have the same number of
successors having exactly $r$ predecessors (for any $r$). But this means that
$n_x$ is invariant. 
\end{proof}
Remark that the same property when removing a node from the bottom layer holds
for the two bottom layers, with symmetric conditions. 

As a consequence of the above proposition and remark, we can get an explicit
expression of these probabilities.
\begin{proposition}\label{prop:3}
Consider the poset $T_H[h,k,|I|]$ and suppose that it is regular. Then the
probabilities $\PP_u(T_H[h,k,|I|])$ that node $x$ of the upper layer terminates
a linear extension, and $\PP_l(T_H[h,k,|I|])$ that isolated node $y$ of the
lower layer terminates a linear extension are given by
\begin{align}
  \PP_u(T_H[h,k,|I|]) & = \frac{1}{h}\frac{\prod_{i=1}^{|I'|}(h-1+k-|I'|+i)}{
      \prod_{i=1}^{|I'|}(h-1+k-|I'|+i) + |I|\times\prod_{i=1}^{|I''|}(h-1+k-|I'|+i)\prod_{i=1}^{|I|-1}(h+k-|I|+i)}\label{eq:pu}\\
  \PP_l(T_H[h,k,|I|]) & = \frac{\prod_{i=1}^{|I''|}(h-1+k-|I'|+i)\prod_{i=1}^{|I|-1}(h+k-|I|+i)}{\prod_{i=1}^{|I'|}(h-1+k-|I'|+i) + |I|\times\prod_{i=1}^{|I''|}(h-1+k-|I'|+i)\prod_{i=1}^{|I|-1}(h+k-|I|+i)}\label{eq:pl},
\end{align}
where $I'$ is the set of isolated nodes in the poset $T_{H\setminus\{x\}}$, and
$I\cup I''=I'$. 
\end{proposition}
\begin{proof}
Removing a node $x$ of the upper layer of $T_H[h,k,|I|]$ yields the poset
$T_{H\setminus\{x\}}[h-1,k,|I'|]$ with $I'\supseteq I$. Then by Lemma~\ref{lem:1}
\begin{equation}\label{eq:11}
e(T_{H\setminus \{x\}}[h-1,k,|I'|]) = e(T_{H\setminus(\{x\}\cup I' )}[h-1,k-|I'|,0])\times\prod_{i=1}^{|I'|}(h-1+k-|I'|+i).
\end{equation}
Removing an isolated node $y$ of the lower layer yields the poset $T_{H\setminus
  \{y\}}[h,k-1,|I|-1]$. Removing all isolated nodes and one node $x$ of the
upper layer yields the poset $T_{H\setminus(I\cup\{x\})}[h-1,k-|I|,|I''|]$, with
the disjoint union $I\cup I''=I'$, i.e., $|I|+|I''|=|I'|$. Applying again
Lemma~\ref{lem:1} we obtain
\begin{align}
e(T_{H\setminus \{y\}}[h,k-1,|I|-1] & = e(T_{H\setminus
  I}[h,k-|I|,0]\prod_{i=1}^{|I|-1}(h+k-|I|+i)\nonumber\\
& = h\times
e(T_{H\setminus(I\cup\{x\})}[h-1,k-|I|,|I''|]\prod_{i=1}^{|I|-1}(h+k-|I|+i)\nonumber\\
 &= h\times e(T_{H\setminus(I'\cup\{x\})}[h-1,k-|I'|,0]\prod_{i=1}^{|I''|}(h-1+k-|I'|+i)\prod_{i=1}^{|I|-1}(h+k-|I|+i)\label{eq:12}
\end{align}
where in the second equality we have used the fact that $T_{H\setminus
  I}[h,k-|I|,0]$ is regular, as well as
Proposition~\ref{prop:1}.

By (\ref{eq:3}), we have for any node $x$ of the upper level and any isolated
node $y$:
\begin{align*}
  \PP_u(T_H[h,k,|I|]) &
  =\frac{e(T_{H\setminus\{x\}}[h-1,k,|I'|])}{e(T_H[h,k,|I|])}\\
&  = \frac{e(T_{H\setminus\{x\}}[h-1,k,|I'|])}{h\times e(T_{H\setminus
      \{x\}}[h-1,k,|I'|]) + |I|\times e(T_{H\setminus\{y\}}[h,k-1,|I|-1])}\\
    & = \frac{1}{h}\frac{\prod_{i=1}^{|I'|}(h-1+k-|I'|+i)}{
      \prod_{i=1}^{|I'|}(h-1+k-|I'|+i) + |I|\times\prod_{i=1}^{|I''|}(h-1+k-|I'|+i)\prod_{i=1}^{|I|-1}(h+k-|I|+i)}
\end{align*}
where in the second equality we have used Proposition~\ref{prop:2}, and in the
third equality Eqs. (\ref{eq:11}) and (\ref{eq:12}). Similarly, we get
  \begin{align*}
  \PP_l(T_H[h,k,|I|]) &
  =\frac{e(T_{H\setminus\{y\}}[h,k-1,|I|-1])}{h\times e(T_{H\setminus
      \{x\}}[h-1,k,|I'|]) + |I|\times e(T_{H\setminus\{y\}}[h,k-1,|I|-1])}\\
     & = \frac{\prod_{i=1}^{|I''|}(h-1+k-|I'|+i)\prod_{i=1}^{|I|-1}(h+k-|I|+i)}{\prod_{i=1}^{|I'|}(h-1+k-|I'|+i) + |I|\times\prod_{i=1}^{|I''|}(h-1+k-|I'|+i)\prod_{i=1}^{|I|-1}(h+k-|I|+i)}.
  \end{align*}

\end{proof}
As a conclusion,
\begin{equation}\label{eq:13}
  \Pr(M\mid T_H[h,k,|I|]) = \begin{cases}
    \PP_u(T_H[h,k,|I|]), & \text{ if } M\text{ is a node of the upper layer}\\
    \PP_l(T_H[h,k,|I|]), & \text{ if } M\text{ is an isolated  node of the lower layer.}
    \end{cases}
\end{equation}
Similar expressions can be obtained for $\Pr(m\mid B_H(h',k',|I'|))$. Both can
be used in Algorithm {\bf generate-linext}. Note that these expressions of
probabilities are approximative, as the assumption that both $ T_H(h,k,|I|)$ and
$ B_H(h',k',|I'|)$ are regular are valid in the first iteration, but not
necessarily in the subsequent iterations.

\subsection{Example}
\begin{figure}[p]
\begin{minipage}{0.6\linewidth}
\begin{tikzpicture}[scale = 1.5]
\node[right] at (-1.1,1-0.05){$1$};\node[right] at (0-0.1,1-0.05){$2$};
\node[right] at (1-0.1,1-0.05){$3$};\node[right] at (2-0.1,1-0.05){$4$};
\node[right] at (-2-0.1,2-0.05){$12$};\node[right] at (-1-0.1,2-0.05){$13$};
\node[right] at (0-0.1,2-0.05){$14$};\node[right] at (1-0.3,2-0.05){$23$};
\node[right] at (3-0.3,2-0.05){$34$};\node[right] at (2-0.1,2-0.05){$24$};
\node[right] at (0.1,3-0.25){$124$}; \node[right] at (2+0.1,3-0.25){$234$};
\node[right] at (-1.1,3-0.25){$123$};\node[right] at (1,3-0.25){$134$};
\draw[-=0.5](0,1-0.25) -- (-2,2-0.25);\draw[fill] (0.0,0.75) circle (0.04);();
\draw[-=0.5](0,1-0.25) -- (2,2-0.25);\draw[-=0.5](0,1-0.25) -- (1,2-0.25);
\draw[-=0.5](-1,1-0.25) -- (-2,2-0.25);\draw[-=0.5](-1,1-0.25) -- (-1,2-0.25);
\draw[-=0.5](-1,1-0.25) -- (0,2-0.25);\draw[fill] (-1,0.75) circle (0.04);();
\draw[-=0.5](1,1-0.25) -- (-1,2-0.25);\draw[-=0.5](1,1-0.25) -- (1,2-0.25);
\draw[-=0.5](1,1-0.25) -- (3,2-0.25);\draw[fill] (1,0.75) circle (0.04);();
\draw[-=0.5](2,1-0.25) -- (3,2-0.25);\draw[-=0.5](2,1-0.25) -- (2,2-0.25);\draw[-=0.5](2,1-0.25) -- (0,2-0.25);
\draw[fill] (2,0.75) circle (0.04);();\draw[fill] (-1,1.75) circle (0.04);();
\draw[color = red,fill] (-1,2.75) circle (0.04);();\draw[color = red,fill] (1,2.75) circle (0.04);();
\draw[fill] (0,1.75) circle (0.04);();\draw[fill] (-2,1.75) circle (0.04);();
\draw[fill] (1,1.75) circle (0.04);();\draw[fill] (2,1.75) circle (0.04);();
\draw[fill] (3,1.75) circle (0.04);();\draw[color = red,fill] (0,2.75) circle (0.04);
\draw[color = red,fill] (2,2.75) circle (0.04);();
\draw [very thick, color = red, dotted] (-2.1,1.6) rectangle (3.1,2.9);
\draw[-=0.5](-2,2-0.25) -- (-1,3-0.25);\draw[-=0.5](-1,2-0.25) -- (-1,3-0.25);
\draw[-=0.5](1,2-0.25) -- (-1,3-0.25);\draw[-=0.5](-1,2-0.25) -- (1,3-0.25);
\draw[-=0.5](0,2-0.25) -- (1,3-0.25);\draw[-=0.5](3,2-0.25) -- (1,3-0.25);
\draw[-=0.5](-2,2-0.25) -- (0,3-0.25);\draw[-=0.5](2,2-0.25) -- (0,3-0.25);
\draw[-=0.5](2,2-0.25) -- (2,3-0.25);\draw[-=0.5](1,2-0.25) -- (2,3-0.25);
\draw[-=0.5](3,2-0.25) -- (2,3-0.25);\draw[-=0.5](0,2-0.25) -- (0,3-0.25);
\node[right] at (0.25,0.4){$(a)$};
\end{tikzpicture} 
\end{minipage}
\hfill
\begin{minipage}{0.6\linewidth}
\begin{tikzpicture}[scale = 1.5]
\node[right] at (-1.1,1-0.05){$1$};\node[right] at (0-0.1,1-0.05){$2$};
\node[right] at (1-0.1,1-0.05){$3$};\node[right] at (2-0.1,1-0.05){$4$};
\node[right] at (-2-0.1,2-0.05){$12$};\node[right] at (-1-0.1,2-0.05){$13$};
\node[right] at (0-0.1,2-0.05){$14$};\node[right] at (1-0.3,2-0.05){$23$};
\node[right] at (3-0.3,2-0.05){$34$};\node[right] at (2-0.1,2-0.05){$24$};
\node[right] at (0.1,3-0.25){$124$};
\node[right] at (-1.1,3-0.25){$123$};\node[right] at (1,3-0.25){$134$};
\draw[-=0.5](0,1-0.25) -- (-2,2-0.25);\draw[color = green,fill] (0.0,0.75) circle (0.04);();
\draw[-=0.5](0,1-0.25) -- (2,2-0.25);\draw[-=0.5](0,1-0.25) -- (1,2-0.25);
\draw[-=0.5](-1,1-0.25) -- (-2,2-0.25);\draw[-=0.5](-1,1-0.25) -- (-1,2-0.25);
\draw[-=0.5](-1,1-0.25) -- (0,2-0.25);\draw[color = green,fill] (-1,0.75) circle (0.04);();
\draw[-=0.5](1,1-0.25) -- (-1,2-0.25);\draw[-=0.5](1,1-0.25) -- (1,2-0.25);
\draw[-=0.5](1,1-0.25) -- (3,2-0.25);\draw[color = green,fill] (1,0.75) circle (0.04);();
\draw[-=0.5](2,1-0.25) -- (3,2-0.25);\draw[-=0.5](2,1-0.25) -- (2,2-0.25);\draw[-=0.5](2,1-0.25) -- (0,2-0.25);
\draw[color = green,fill] (2,0.75) circle (0.04);();\draw[fill] (-1,1.75) circle (0.04);();
\draw[fill] (-1,2.75) circle (0.04);();\draw[fill] (1,2.75) circle (0.04);();
\draw[fill] (0,1.75) circle (0.04);();\draw[fill] (-2,1.75) circle (0.04);();
\draw[color = white,fill] (0,0) circle (0.04);
\draw[fill] (1,1.75) circle (0.04);();\draw[fill] (2,1.75) circle (0.04);();
\draw[fill] (3,1.75) circle (0.04);();\draw[fill] (0,2.75) circle (0.04);();
\draw [very thick, color = green, dotted] (-2.2,0.6) rectangle (3.3,1.9);
\draw[-=0.5](-2,2-0.25) -- (-1,3-0.25);\draw[-=0.5](-1,2-0.25) -- (-1,3-0.25);
\draw[-=0.5](1,2-0.25) -- (-1,3-0.25);\draw[-=0.5](-1,2-0.25) -- (1,3-0.25);
\draw[-=0.5](0,2-0.25) -- (1,3-0.25);\draw[-=0.5](3,2-0.25) -- (1,3-0.25);
\draw[-=0.5](-2,2-0.25) -- (0,3-0.25);\draw[-=0.5](2,2-0.25) -- (0,3-0.25);
\draw[-=0.5](0,2-0.25) -- (0,3-0.25);
\node[right] at (0.25,0.4){$(b)$};
\end{tikzpicture} 
\end{minipage}
\hfill
\begin{minipage}{0.6\linewidth}
\begin{tikzpicture}[scale = 1.5]
\node[right] at (0-0.1,1-0.05){$2$};
\node[right] at (1-0.1,1-0.05){$3$};\node[right] at (2-0.1,1-0.05){$4$};
\node[right] at (-2-0.1,2-0.05){$12$};\node[right] at (-1-0.1,2-0.05){$13$};
\node[right] at (0-0.1,2-0.05){$14$};\node[right] at (1-0.3,2-0.05){$23$};
\node[right] at (3-0.3,2-0.05){$34$};\node[right] at (2-0.1,2-0.05){$24$};
\node[right] at (0.1,3-0.25){$124$};\draw[fill] (2.0,0.75) circle (0.04);
\node[right] at (-1.1,3-0.25){$123$};\node[right] at (1,3-0.25){$134$};
\draw[-=0.5](0,1-0.25) -- (-2,2-0.25);\draw[fill] (0.0,0.75) circle (0.04);
\draw[fill] (-1,1.75) circle (0.04);();
\draw[-=0.5](0,1-0.25) -- (2,2-0.25);\draw[-=0.5](0,1-0.25) -- (1,2-0.25);
\draw[-=0.5](1,1-0.25) -- (-1,2-0.25);\draw[-=0.5](1,1-0.25) -- (1,2-0.25);
\draw[-=0.5](1,1-0.25) -- (3,2-0.25);\draw[fill] (1,0.75) circle (0.04);();
\draw[-=0.5](2,1-0.25) -- (3,2-0.25);\draw[-=0.5](2,1-0.25) -- (2,2-0.25);\draw[-=0.5](2,1-0.25) -- (0,2-0.25);
\draw [very thick, color = red, dotted] (-2.1,1.6) rectangle (3.1,2.9);
\draw[color = red,fill] (-1,2.75) circle (0.04);();\draw[color = red,fill] (1,2.75) circle (0.04);();
\draw[fill] (0,1.75) circle (0.04);();\draw[fill] (-2,1.75) circle (0.04);();
\draw[fill] (1,1.75) circle (0.04);();\draw[fill] (2,1.75) circle (0.04);();
\draw[fill] (3,1.75) circle (0.04);();\draw[color = red,fill] (0,2.75) circle (0.04);();
\draw[-=0.5](-2,2-0.25) -- (-1,3-0.25);\draw[-=0.5](-1,2-0.25) -- (-1,3-0.25);
\draw[-=0.5](1,2-0.25) -- (-1,3-0.25);\draw[-=0.5](-1,2-0.25) -- (1,3-0.25);
\draw[-=0.5](0,2-0.25) -- (1,3-0.25);\draw[-=0.5](3,2-0.25) -- (1,3-0.25);
\draw[-=0.5](-2,2-0.25) -- (0,3-0.25);\draw[-=0.5](2,2-0.25) -- (0,3-0.25);
\draw[-=0.5](0,2-0.25) -- (0,3-0.25);
\node[right] at (0.25,0.4){$(c)$};
\end{tikzpicture} 
\end{minipage}
\hfill
\begin{minipage}{0.6\linewidth}
\begin{tikzpicture}[scale = 1.5]
\node[right] at (0-0.1,1-0.05){$2$};
\node[right] at (1-0.1,1-0.05){$3$};\node[right] at (2-0.1,1-0.05){$4$};
\node[right] at (-2-0.1,2-0.05){$12$};\node[right] at (-1-0.1,2-0.05){$13$};
\node[right] at (0-0.1,2-0.05){$14$};\node[right] at (1-0.3,2-0.05){$23$};
\node[right] at (3-0.3,2-0.05){$34$};\node[right] at (2-0.1,2-0.05){$24$};
\node[right] at (0.1,3-0.25){$124$};\node[right] at (-1.1,3-0.25){$123$};
\draw[-=0.5](0,1-0.25) -- (-2,2-0.25);\draw[color = green,fill] (0.0,0.75) circle (0.04);();
\draw[-=0.5](0,1-0.25) -- (2,2-0.25);\draw[-=0.5](0,1-0.25) -- (1,2-0.25);
\draw[-=0.5](1,1-0.25) -- (-1,2-0.25);\draw[-=0.5](1,1-0.25) -- (1,2-0.25);
\draw[-=0.5](1,1-0.25) -- (3,2-0.25);\draw[color = green,fill] (1,0.75) circle (0.04);();
\draw[-=0.5](2,1-0.25) -- (3,2-0.25);\draw[-=0.5](2,1-0.25) -- (2,2-0.25);\draw[-=0.5](2,1-0.25) -- (0,2-0.25);
\draw[color = green,fill] (2,0.75) circle (0.04);();\draw[fill] (-1,1.75) circle (0.04);();
\draw[fill] (-1,2.75) circle (0.04);();\draw[fill] (0,1.75) circle (0.04);();\draw[fill] (-2,1.75) circle (0.04);();
\draw[fill] (1,1.75) circle (0.04);();\draw[fill] (2,1.75) circle (0.04);();
\draw[fill] (3,1.75) circle (0.04);();\draw[fill] (0,2.75) circle (0.04);();
\draw [very thick, color = green, dotted] (-2.2,0.6) rectangle (3.3,1.9);
\draw[-=0.5](-2,2-0.25) -- (-1,3-0.25);\draw[-=0.5](-1,2-0.25) -- (-1,3-0.25);
\draw[-=0.5](1,2-0.25) -- (-1,3-0.25);\draw[-=0.5](-2,2-0.25) -- (0,3-0.25);\draw[-=0.5](2,2-0.25) -- (0,3-0.25);\draw[color = white,fill] (0,0) circle (0.04);
\draw[-=0.5](0,2-0.25) -- (0,3-0.25);\node[right] at (0.25,0.4){$(d)$};
\end{tikzpicture} 
\end{minipage}
\begin{minipage}{0.6\linewidth}
\begin{tikzpicture}[scale = 1.5]
\node[right] at (0-0.1,1-0.05){$2$};
\node[right] at (1-0.1,1-0.05){$3$};
\node[right] at (-2-0.1,2-0.05){$12$};\node[right] at (-1-0.1,2-0.05){$13$};
\node[right] at (0-0.1,2-0.05){$14$};\node[right] at (1-0.3,2-0.05){$23$};
\node[right] at (3-0.3,2-0.05){$34$};\node[right] at (2-0.1,2-0.05){$24$};
\node[right] at (0.1,3-0.25){$124$};\node[right] at (-1.1,3-0.25){$123$};
\draw[-=0.5](0,1-0.25) -- (-2,2-0.25);\draw[fill] (0.0,0.75) circle (0.04);();
\draw[-=0.5](0,1-0.25) -- (2,2-0.25);\draw[-=0.5](0,1-0.25) -- (1,2-0.25);
\draw[-=0.5](1,1-0.25) -- (-1,2-0.25);\draw[-=0.5](1,1-0.25) -- (1,2-0.25);
\draw[-=0.5](1,1-0.25) -- (3,2-0.25);\draw[fill] (1,0.75) circle (0.04);();
\draw[fill] (-1,1.75) circle (0.04);();
\draw[color=red,fill] (-1,2.75) circle (0.04);();\draw[fill] (0,1.75) circle (0.04);();\draw[fill] (-2,1.75) circle (0.04);();
\draw[fill] (1,1.75) circle (0.04);();\draw[fill] (2,1.75) circle (0.04);();
\draw[color = red,fill] (3,1.75) circle (0.04);();\draw[color=red,fill] (0,2.75) circle (0.04);();
\draw [very thick, color = red, dotted] (-2.1,1.6) rectangle (3.1,2.9);
\draw[-=0.5](-2,2-0.25) -- (-1,3-0.25);\draw[-=0.5](-1,2-0.25) -- (-1,3-0.25);
\draw[-=0.5](1,2-0.25) -- (-1,3-0.25);\draw[-=0.5](-2,2-0.25) -- (0,3-0.25);\draw[-=0.5](2,2-0.25) -- (0,3-0.25);
\draw[-=0.5](0,2-0.25) -- (0,3-0.25);\node[right] at (0.25,0.4){$(e)$};
\end{tikzpicture} 
\end{minipage}
\hfill
\begin{minipage}{0.6\linewidth}
\begin{tikzpicture}[scale = 1.5]
\node[right] at (0-0.1,1-0.05){$2$};
\node[right] at (1-0.1,1-0.05){$3$};\node[right] at (-2-0.1,2-0.05){$12$};\node[right] at (-1-0.1,2-0.05){$13$};
\node[right] at (0-0.1,2-0.05){$14$};\node[right] at (1-0.3,2-0.05){$23$};
\node[right] at (3-0.3,2-0.05){$34$};\node[right] at (2-0.1,2-0.05){$24$};
\node[right] at (-1.1,3-0.25){$123$};
\draw[-=0.5](0,1-0.25) -- (-2,2-0.25);\draw[color = green,fill] (0.0,0.75) circle (0.04);();
\draw[-=0.5](0,1-0.25) -- (2,2-0.25);\draw[-=0.5](0,1-0.25) -- (1,2-0.25);
\draw[-=0.5](1,1-0.25) -- (-1,2-0.25);\draw[-=0.5](1,1-0.25) -- (1,2-0.25);
\draw[-=0.5](1,1-0.25) -- (3,2-0.25);\draw[color = green,fill] (1,0.75) circle (0.04);();
\draw[fill] (-1,1.75) circle (0.04);();
\draw[fill] (-1,2.75) circle (0.04);();\draw[color=green,fill] (0,1.75) circle (0.04);();\draw[fill] (-2,1.75) circle (0.04);();
\draw[fill] (1,1.75) circle (0.04);();\draw[fill] (2,1.75) circle (0.04);();
\draw[fill] (3,1.75) circle (0.04);();
\draw [very thick, color = green, dotted] (-2.2,0.6) rectangle (3.3,1.9);
\draw[-=0.5](-2,2-0.25) -- (-1,3-0.25);\draw[-=0.5](-1,2-0.25) -- (-1,3-0.25);
\draw[-=0.5](1,2-0.25) -- (-1,3-0.25);\draw[color = white,fill] (0,0) circle (0.04);\node[right] at (0.25,0.4){$(f)$};
\end{tikzpicture} 
\end{minipage}

\begin{minipage}{0.6\linewidth}
\begin{tikzpicture}[scale = 1.5]
\node[right] at (0-0.1,1-0.05){$2$};
\node[right] at (1-0.1,1-0.05){$3$};\node[right] at (-2-0.1,2-0.05){$12$};\node[right] at (-1-0.1,2-0.05){$13$};\node[right] at (1-0.3,2-0.05){$23$};
\node[right] at (3-0.3,2-0.05){$34$};\node[right] at (2-0.1,2-0.05){$24$};
\node[right] at (-1.1,3-0.25){$123$};
\draw[-=0.5](0,1-0.25) -- (-2,2-0.25);\draw[fill] (0.0,0.75) circle (0.04);();
\draw[-=0.5](0,1-0.25) -- (2,2-0.25);\draw[-=0.5](0,1-0.25) -- (1,2-0.25);
\draw[-=0.5](1,1-0.25) -- (-1,2-0.25);\draw[-=0.5](1,1-0.25) -- (1,2-0.25);
\draw[-=0.5](1,1-0.25) -- (3,2-0.25);\draw[fill] (1,0.75) circle (0.04);();
\draw[fill] (-1,1.75) circle (0.04);();
\draw[color = red,fill] (-1,2.75) circle (0.04);();\draw[fill] (-2,1.75) circle (0.04);();
\draw[fill] (1,1.75) circle (0.04);();\draw[color = red,fill] (2,1.75) circle (0.04);();
\draw[color = red,fill] (3,1.75) circle (0.04);();
\draw [very thick, color = red, dotted] (-2.1,1.6) rectangle (3.1,2.9);
\draw[-=0.5](-2,2-0.25) -- (-1,3-0.25);\draw[-=0.5](-1,2-0.25) -- (-1,3-0.25);
\draw[-=0.5](1,2-0.25) -- (-1,3-0.25);\draw[color = white,fill] (0,0) circle (0.04);\node[right] at (0.25,0.4){$(g)$};
\end{tikzpicture} 
\end{minipage}
\hfill
\begin{minipage}{0.6\linewidth}
\begin{tikzpicture}[scale = 1.5]
\node[right] at (0-0.1,1-0.05){$2$};
\node[right] at (1-0.1,1-0.05){$3$};\node[right] at (-2-0.1,2-0.05){$12$};\node[right] at (-1-0.1,2-0.05){$13$};
\node[right] at (1-0.3,2-0.05){$23$};
\node[right] at (3-0.3,2-0.05){$34$};\node[right] at (2-0.1,2-0.05){$24$};
\draw[-=0.5](1,1-0.25) -- (-1,2-0.25);\draw[-=0.5](1,1-0.25) -- (1,2-0.25);
\draw[-=0.5](0,1-0.25) -- (-2,2-0.25);\draw[color = green,fill] (0.0,0.75) circle (0.04);();
\draw[-=0.5](0,1-0.25) -- (2,2-0.25);\draw[-=0.5](0,1-0.25) -- (1,2-0.25);
\draw[-=0.5](1,1-0.25) -- (3,2-0.25);\draw[color = green,fill] (1,0.75) circle (0.04);();
\draw[fill] (-1,1.75) circle (0.04);\draw[fill] (-2,1.75) circle (0.04);
\draw[fill] (1,1.75) circle (0.04);\draw[fill] (2,1.75) circle (0.04);
\draw[fill] (3,1.75) circle (0.04);
\draw [very thick, color = green, dotted] (-2.2,0.6) rectangle (3.3,1.9);\node[right] at (0.25,0.4){$(h)$};
\end{tikzpicture} 
\end{minipage}
\hfill
\begin{minipage}{0.6\linewidth}
\begin{tikzpicture}[scale = 1.5]
\node[right] at (1-0.1,1-0.05){$3$};\node[right] at (-2-0.1,2-0.05){$12$};\node[right] at (-1-0.1,2-0.05){$13$};
\node[right] at (1-0.3,2-0.05){$23$};
\node[right] at (3-0.3,2-0.05){$34$};\node[right] at (2-0.1,2-0.05){$24$};
\draw[-=0.5](1,1-0.25) -- (-1,2-0.25);\draw[-=0.5](1,1-0.25) -- (1,2-0.25);
\draw[-=0.5](1,1-0.25) -- (3,2-0.25);\draw[color = green,fill] (1,0.75) circle (0.04);();
\draw[fill] (-1,1.75) circle (0.04);();\draw[color = green,fill] (-2,1.75) circle (0.04);();
\draw[fill] (1,1.75) circle (0.04);();\draw[color = green,fill] (2,1.75) circle (0.04);();
\draw[fill] (3,1.75) circle (0.04);();
\draw [very thick, color = green, dotted] (-2.2,0.6) rectangle (3.3,1.9);\node[right] at (0.25,0.4){$(i)$};
\end{tikzpicture} 
\end{minipage}
\hfill
\begin{minipage}{0.6\linewidth}
\begin{tikzpicture}[scale = 1.5]
\node[right] at (-2-0.1,2-0.05){$12$};\node[right] at (-1-0.1,2-0.05){$13$};
\node[right] at (1-0.3,2-0.05){$23$};
\node[right] at (3-0.3,2-0.05){$34$};\node[right] at (2-0.1,2-0.05){$24$};
\draw[color = green,fill] (-1,1.75) circle (0.04);();\draw[color = green,fill] (-2,1.75) circle (0.04);();
\draw[color = green,fill] (1,1.75) circle (0.04);();\draw[color = green,fill] (2,1.75) circle (0.04);();
\draw[color = green,fill] (3,1.75) circle (0.04);();
\draw [very thick, color = green, dotted] (-2.2,1.2) rectangle (3.3,1.9);\node[right] at (0.25,0.4){$(j)$};
\end{tikzpicture} 
\end{minipage}
\caption{Process of node selection according to Algorithm {\bf
    generate-linext}. The dotted red and green boxes indicate the two top and
  bottom layers under consideration, respectively. Red nodes are maximal
  elements in the two top layers, while green nodes are minimal elements in the
  two bottom layers.}
\label{fig:ill}
\end{figure}
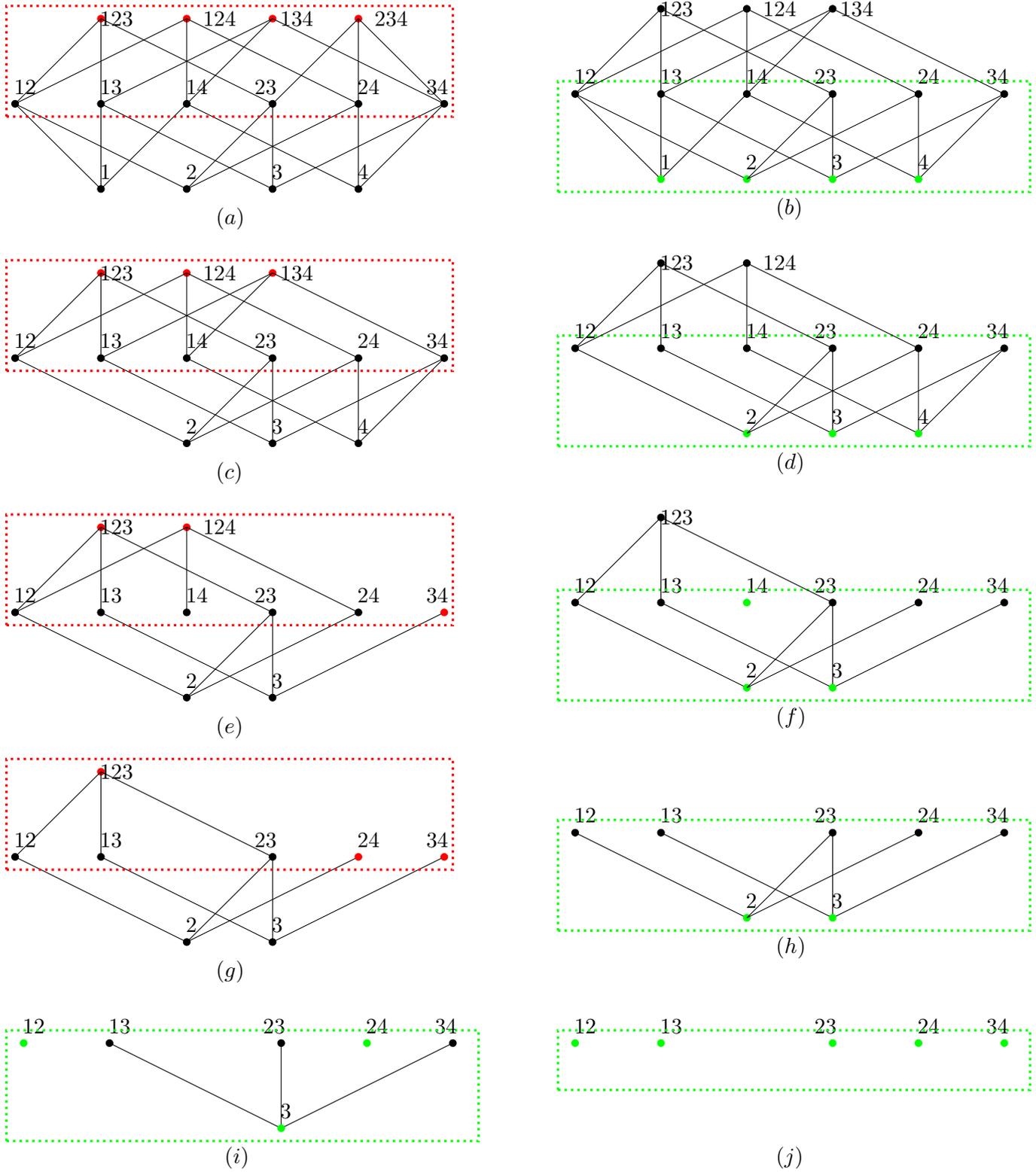

We illustrate the 2-layer approximation algorithm by applying it for generating a linear
extension with $n=4$ (see Figure~\ref{fig:ill} (a), (b), ... (i)).
For simplicity, sets are denoted by 123 instead of $\{1,2,3\}$, etc.
\begin{enumerate}
\item  (a): the two top layers are considered. Here, since
  all maximal nodes are on the first layer, by Proposition~\ref{prop:2}, all
  maximal nodes have the same probability of 0.25. Suppose 234 is
  selected, which yields Figure (b).
\item (b): now the two bottom layers are
  considered. Similarly, all minimal nodes
are equiprobable. Suppose node 1 is selected, this yields (c). 
\item (c): As for (a), we consider again the two top layers. All maximal nodes are on the first layer, hence
  equiprobable. Supposing 134 is selected, (d) is obtained. 
\item (d): As for (b), we consider the two bottom layers, whose minimal nodes
  are on the bottom layer, hence equiprobable. Supposing node 4 is selected, we
  obtain (e).
\item (e): In the two upper layers, one maximal node is on the second
  layer. Then Proposition~\ref{prop:3} has to be used to compute the
  probabilities for the first and second layer, which are respectively 0.437 and
  0.125. Supposing 124 is selected, (f) is obtained.
\item (f): In the two bottom layers, one minimal element is 14 on
  the second layer, hence again Proposition~\ref{prop:3} must be used. One finds
  0.437 for  nodes 2 and 3,    and 0.125 for 14. Supposing
  14 is selected, we obtain (g). 
\item (g): Now the probability of selecting 123 is 0.67, and is 0.167 for
  selecting 24 and 34. Supposing 123 is selected, we obtain (h). 
\item (h): The two top and bottom layers coincide. We consider them as the two
  bottom layers, since there are less minimal elements than maximal elements,
  and select a minimal element. They are equiprobable. Supposing node 2 is
  selected, (i) is obtained.
\item (i): Again, we select a minimal node. Applying Proposition~\ref{prop:3},
  the probabilities are 0.67 for selecting node 3, and 0.167 for selecting 12
  and 24. Supposing node 3 is selected, we obtain (j).
\item (j): there is only one layer, i.e., all nodes are both maximal and
  minimal and are equiprobable. We may suppose that they are selected in this
  order: 12, 34, 24, 23, 13.
\end{enumerate}
Finally, the obtained linear extension is: 1, 4, 14, 2, 3, 12, 34, 24, 23, 13,
123, 124, 134, 234.

\section{Measures of performance}\label{sec:perf}
An important point is to be able to assess the performance of a given algorithm
generating capacities in a uniform way, i.e., how much uniform is the
distribution obtained? Given the high dimension of the polyhedron of capacities,
no graphical view is possible, and the topic appears to be more difficult than
it looks at first sight. In what follows, we propose two ways of measuring the
performance, based on the distribution of $\mu(S)$ for a given subset $S$, and
on the centroid of $\cC(N)$.

\subsection{Distribution of $\mu(S)$}

Uniform distribution of the capacity does not mean that the distribution of
$\mu(S)$ for any $S\subseteq N$, $S\neq\varnothing,N$ is uniform. An explicit
and convenient expression for the distribution of $\mu(S)$ seems however very
difficult to obtain. The following can be said, however.

We denote by $\bmu$ the corresponding random variable. Refering to the notation
of Section~\ref{sec:rgline}, consider a linear extension associated to the
permutation $\sigma$ on $2^N\setminus\{\varnothing,N\}$, and $R_\sigma$ the
corresponding simplex in $\cC(N)$. Supposing $\bmu\in R_\sigma$, we have that
$\bmu(S_{\sigma(k)})$ follows the distribution of the $k$th order statistics on
$[0,1]$. It is known that the probability density function $f_{(k)}$ of the
$k$th order statistics on $[0,1]$ when the underlying $2^n-2$ random variables are
i.i.d. and uniform is a Beta distribution:
\[
f_{(k)}(u) = (2^n-2)\binom{2^n-3}{k-1}(1-u)^{2^n-2-k}u^{k-1} = \Beta(k,2^n+k-1)
\]
with $\Beta(r,s)=\frac{1}{B(r,s)}u^{r-1}(1-u)^{s-1}$, and $B(r,s)$ is the Beta
function
\[
B(r,s) = \int_0^1t^{r-1}(1-t)^{s-1}\dd t = \frac{\Gamma(s)\Gamma(r)}{\Gamma(r+s)}
\]
with $\Gamma(r) = \int_0^1t^{r-1}e^{-t}\dd t$ the Gamma function. Denoting by
$\OS_k$ the corresponding cumulative distribution function, it follows that for
any $S\in 2^N\setminus\{\varnothing,N\}$, the distribution $F_{\bmu(S)}(\alpha)$
is given by
\begin{align}
F_{\bmu(S)}(\alpha) &= \Pr(\bmu(S)\leqslant \alpha) = \sum_{\sigma\in
  E(2^N\setminus\{\varnothing,N\})}\Pr(\bmu(S)\leqslant \alpha\mid\bmu\in
R_\sigma)\Pr(\bmu\in R_\sigma)\nonumber\\
  & = \frac{1}{e(2^N)}\sum_{\sigma\in
    E(2^N\setminus\{\varnothing,N\})}\Pr(\bmu(S)\leqslant \alpha\mid\bmu\in
  R_\sigma)\nonumber\\
  & =  \frac{1}{e(2^N)}\sum_{\sigma\in
    E(2^N\setminus\{\varnothing,N\})}\OS_{k(S,\sigma)}(\alpha), \label{eq:dist}
\end{align}
where  $E(2^N\setminus\{\varnothing,N\})$ is the set of permutations
corresponding to linear extensions, and $k(S,\sigma)$ is such that
$S=S_{\sigma(k)}$. 

Based on this formula, the following can be shown.
\begin{lemma}\label{lem:dist}
  Assume $\bmu$ is uniformly distributed and take $\varnothing\neq S,S'\subset N$. Then
  \begin{enumerate}
    \item $\bmu(S)$ and $\bmu(S')$ for $|S|=|S'|$ are identically distributed.
    \item $\bmu(S)$ and $1-\bmu(N\setminus S)$ are  identically distributed.
  \end{enumerate}
\end{lemma}
\begin{proof}
  (i) Consider distinct subsets $S,S'\in 2^N\setminus\{\varnothing,N\}$ such
  that $|S|=|S'|$, and fix a linear extension $\sigma$. Then there exists a
  permutation $\pi$ on $N$ such that $\pi(S)=S'$, and $k(S,\sigma) =
  k(\pi(S),\pi(\sigma))$, where $\pi(S)=\{\pi(i), i\in S\}$ and $\pi(\sigma)$ is
  the linear extension
  $\pi(S_{\sigma(1)}),\ldots,\pi(S_{\sigma(2^n-2)})$. Applying (\ref{eq:dist}) we
  find:
  \begin{align*}
\Pr(\bmu(S)\leqslant \alpha) &= \frac{1}{e(2^N)}\sum_{\sigma\in
    E(2^N\setminus\{\varnothing,N\})}\OS_{k(S,\sigma)}(\alpha)\\
  &=\frac{1}{e(2^N)}\sum_{\pi(\sigma)\in
    E(2^N\setminus\{\varnothing,N\})}\OS_{k(\pi(S),\pi(\sigma))}(\alpha) =
    \Pr(\bmu(S')\leqslant \alpha),
  \end{align*}
  the last equation following from the fact that $\pi$ is a bijection on
  $E(2^N\setminus\{\varnothing,N\})$.  

  (ii) Consider the conjugate capacity $\overline{\bmu}$ defined by
  $\bmu(S)=1-\bmu(N\setminus S)$. Taking a linear extension
  $S_\sigma(1)<S_{\sigma(2)}<\cdots< S_{\sigma(2^n-2)}$, 
  it is easy to see that the sequence
  \[
N\setminus S_{\sigma(2^n-2)},\ldots,N\setminus S_{\sigma(2)}, N\setminus S_{\sigma(1)}
  \]
is also a linear extension $\overline{\sigma}$ with the property
$k(S,\sigma)=2^n-2-k(N\setminus S,\overline{\sigma})$. Using the fact that $\OS_k$ is
distributed as $1-\OS_{2^n-2-k}$, we obtain
  \begin{align*}
    \Pr(\bmu(S)\leqslant \alpha) &=\frac{1}{e(2^N)}\sum_{\sigma\in
    E(2^N\setminus\{\varnothing,N\})}\OS_{k(S,\sigma)}(\alpha)\\
  &=\frac{1}{e(2^N)}\sum_{\overline{\sigma}\in
    E(2^N\setminus\{\varnothing,N\})}\OS_{2^n-2-k(N\setminus
          S,\overline{\sigma})}(\alpha) =
     1-    \frac{1}{e(2^N)}\sum_{\overline{\sigma}\in
    E(2^N\setminus\{\varnothing,N\})}\OS_{k(N\setminus
          S,\overline{\sigma})}(\alpha)\\
   &=1- \Pr(\bmu(N\setminus S)\leqslant \alpha),
  \end{align*}
  the last equation following from the fact that
  $\sigma\mapsto\overline{\sigma}$ is a bijection on
  $E(2^N\setminus\{\varnothing,N\})$.

\end{proof}

Several methods for measuring the uniformity of the distribution of $\bmu$ may
be deduced from the above results. When $n\leqslant 4$, it is possible to
generate all linear extensions and therefore to have an exact generator for
uniform capacities. It suffices then to compare the histograms obtained for
$\bmu(S)$ for the considered method and the exact method. Thanks to
Lemma~\ref{lem:dist}, we may limit to $\lceil\frac{n}{2}\rceil$ comparisons, as
it suffices to take one set of cardinality $1,2,\ldots, \lceil\frac{n}{2}\rceil$
(provided the properties of Lemma~\ref{lem:dist} are satisfied by the method
under consideration). 

\begin{figure}[p]
\begin{center}
\includegraphics[width=15cm]{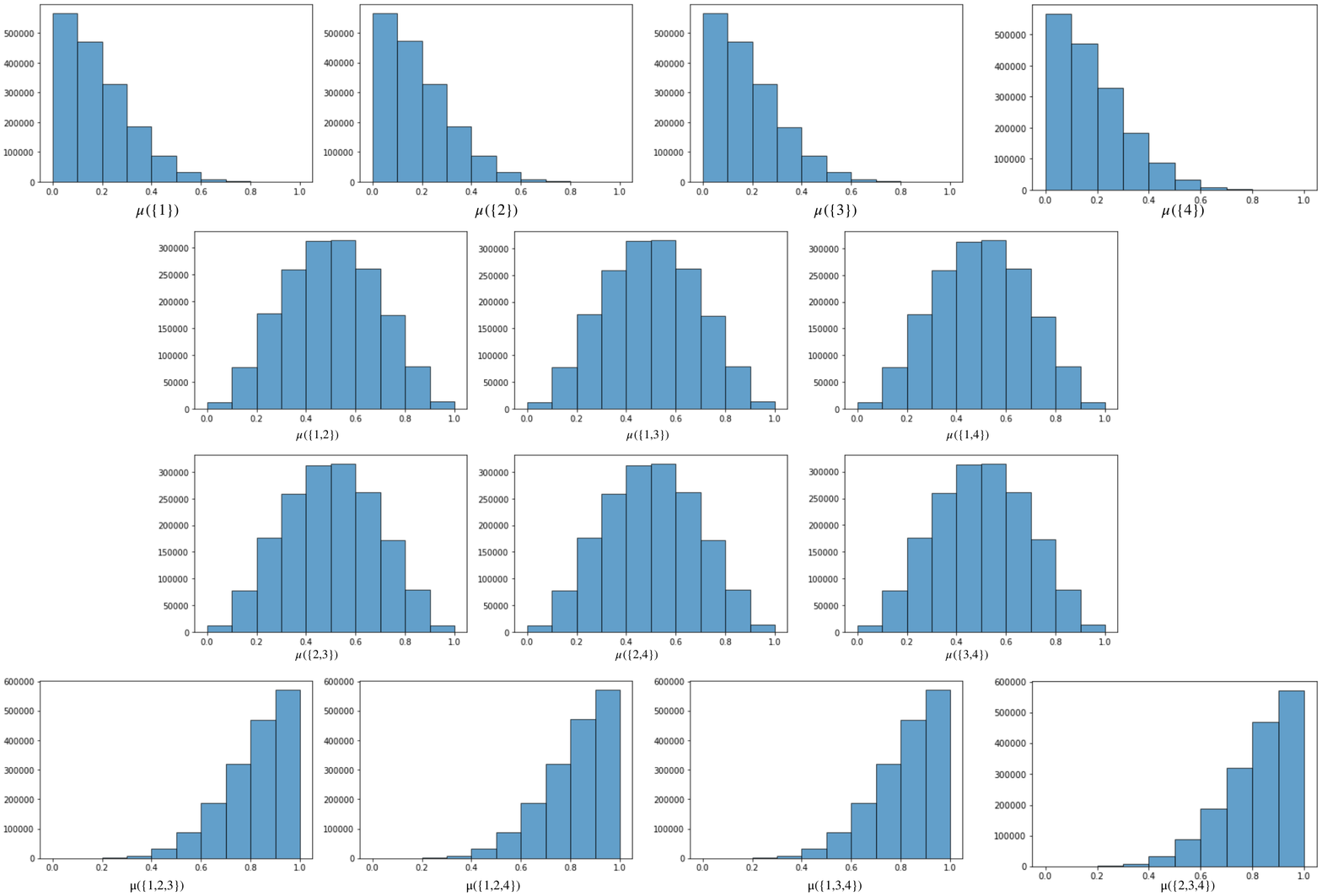}
\end{center}
\caption{Histograms of $\bmu(S)$ for $n=4$ and the exact method}
\label{fig:ex4}
\end{figure}

\begin{figure}[p]
\begin{center}
\includegraphics[width=15cm]{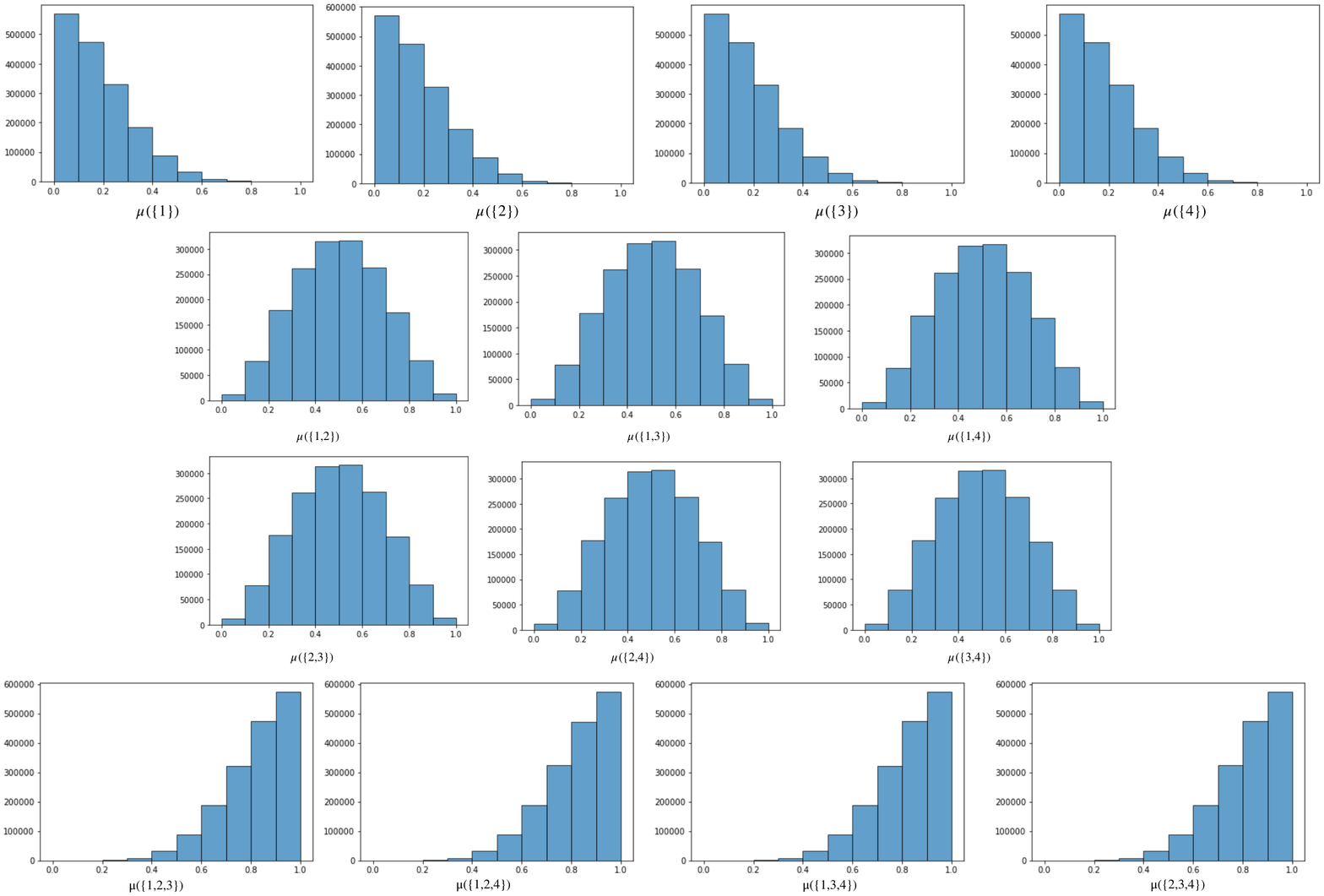}
\end{center}
\caption{Histograms of $\bmu(S)$ for $n=4$ and the 2-layer approximation method}
\label{fig:approx4}
\end{figure}

\begin{figure}[p]
\begin{center}
\includegraphics[width=15cm]{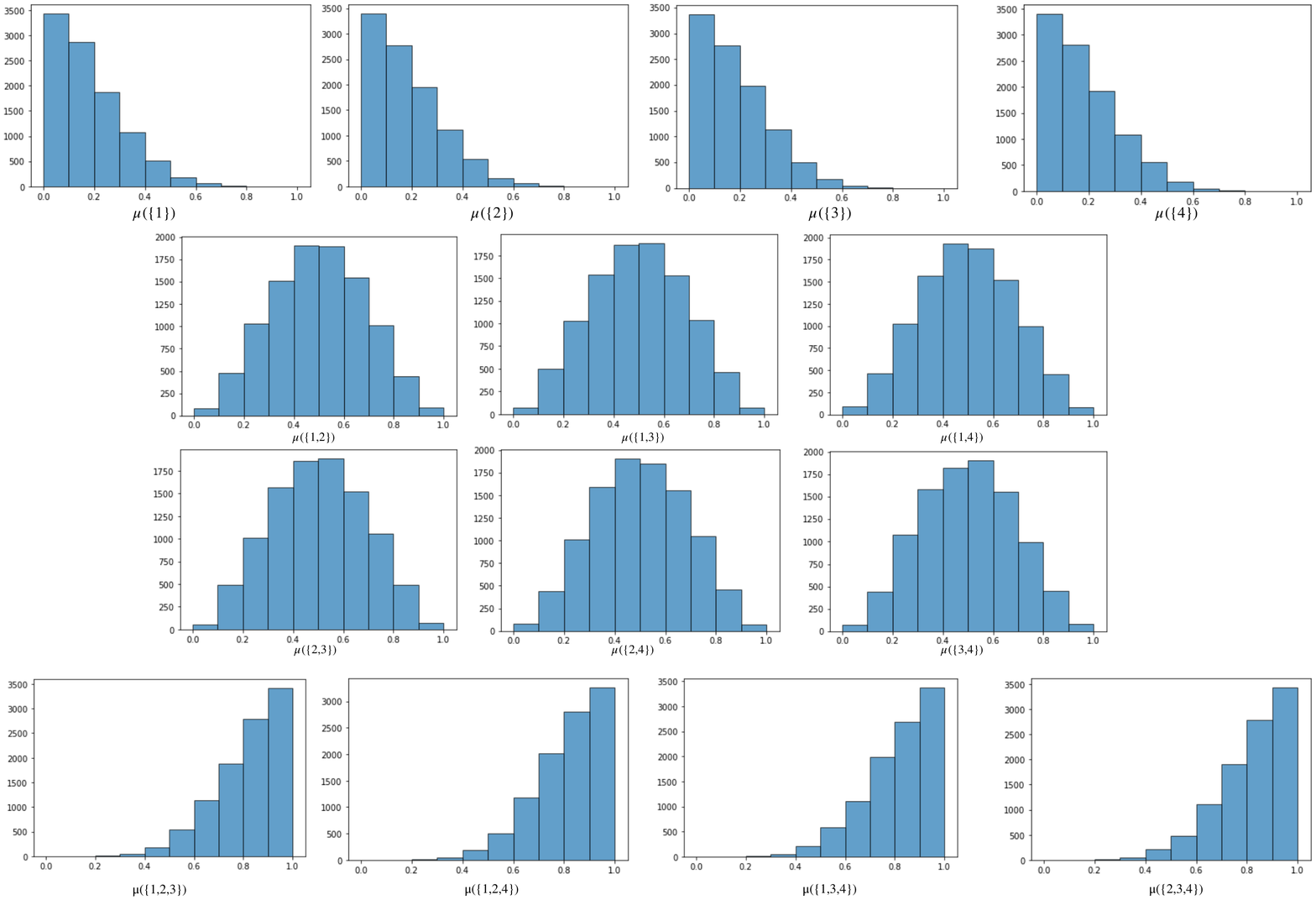}
\end{center}
\caption{Histograms of $\bmu(S)$ for $n=4$ and the Markov chain method}
\label{fig:Markov4}
\end{figure}

\begin{figure}[p]
\begin{center}
\includegraphics[width=15cm]{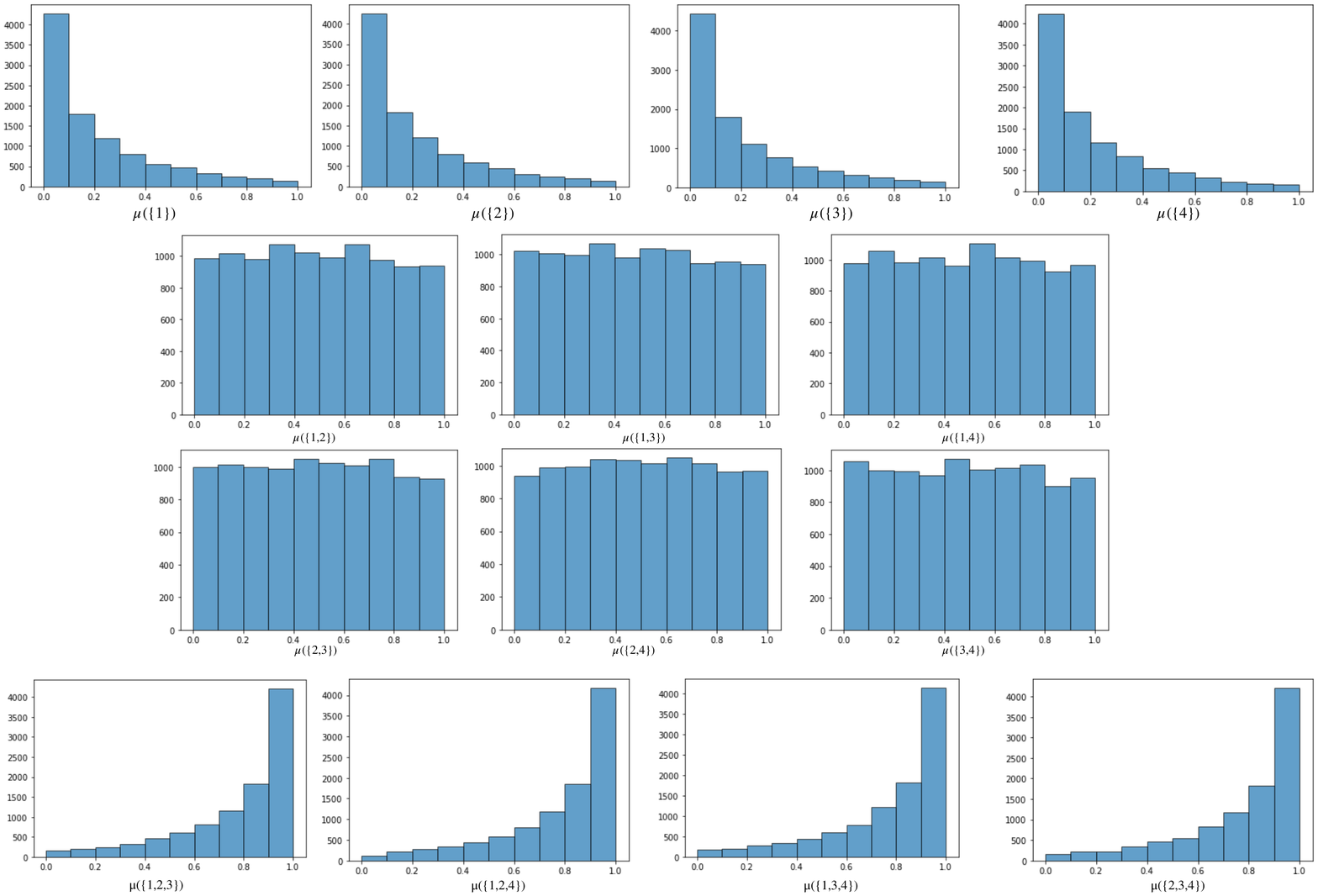}
\end{center}
\caption{Histograms of $\bmu(S)$ for $n=4$ and the Random Node Generator}
\label{fig:rng4}
\end{figure}

Figures~\ref{fig:ex4}, \ref{fig:approx4}, \ref{fig:Markov4} and \ref{fig:rng4}
show the histograms obtained for all subsets $S\in 2^N\setminus{\varnothing,N}$
for $n=4$, respectively for the exact method, the 2-layer approximation method,
the Markov chain generator \cite{kakh91}, and the random node generator \cite{hapi17}.

For comparing the histograms produced by the different methods with those of the
exact methods, we use the Kullback-Leibler divergence. Given two discrete
probability distribution $p,q$ on the same universe $X$, the Kullback-Leibler
diveregence is defined as
\[
\mathbb{D}_{KL}(p||q)= \sum_{x\in X}p(x)\log\frac{p(x)}{q(x)}.
\]
The smaller the value, the closer are the two distributions. We have computed
$\mathbb{D}_{KL}(p||q)$, replacing $p$ with the exact distribution and $q$ with
the distribution obtained by the Markov chain method and the 2-layer
approximation, respectively.  Table~\ref{tab:kl} show the results for $n=4$.

\begin{table}[htb]
  \begin{center}
\begin{tabular}{|c|cccc|}
\hline capacity generator & $\mu(\{1\})$ & $\mu(\{2\})$ &$\mu(\{3\})$ &
    $\mu(\{4\})$ \\ \hline Markov chain &0.6658 & 0.6698 & 0.6597 &
      0.6650\\ 2-layer approximation & 0.6633 & 0.6614 & 0.6637 & 0.6635\\

\hline
\end{tabular}

\medskip

\begin{tabular}{|c|cccccc|}
\hline capacity generator & $\mu(\{1,2\})$ & $\mu(\{1,3\})$ &$\mu(\{1,4\})$ &
      $\mu(\{2,3\})$ & $\mu(\{2,4\})$ & $\mu(\{3,4\})$ \\ \hline Markov chain
            & 0.1595 & 0.1593 & 0.1595 & 0.1593& 0.1596 & 0.1593\\ 2-layer
            approximation & 0.1585 & 0.1586 & 0.1587 & 0.1586& 0.1587 & 0.1586\\

\hline
\end{tabular}

\medskip

\begin{tabular}{|c|cccc|}
\hline capacity generator & $\mu(\{1,2,3\})$ & $\mu(\{1,2,4\})$
    &$\mu(\{1,3,4\})$ & $\mu(\{2,3,4\})$ \\ \hline Markov chain & 0.0303 &
        0.0304 & 0.0303 & 0.0303\\ 2-layer approximation & 0.0299 & 0.0300 &
        0.0299 & 0.0299\\

\hline
\end{tabular}
  \end{center}
  \caption{Kullback-Leibler divergence for the 2-layer approximation method and
    the Markov chain method compared to the exact distribution ($n=4$)}
  \label{tab:kl}
\end{table}
As it can be seen, the divergences obtained for the two methods are very close,
with a slight advantage for the 2-layer approximation method since its values
are systematically smaller.

\medskip

\begin{figure}[p]
\begin{center}
\includegraphics[width=10cm]{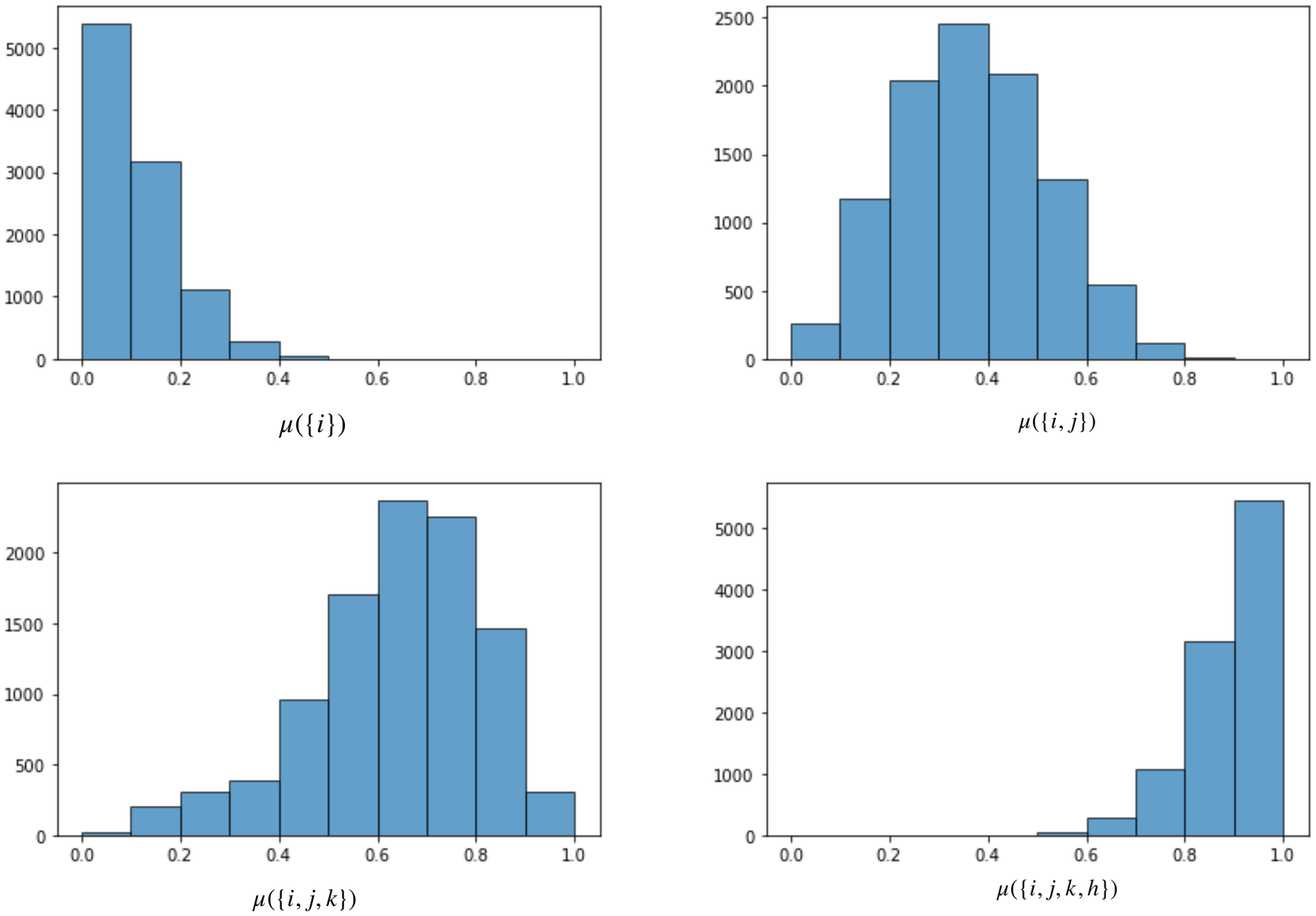}
\end{center}
\caption{Histograms of $\bmu(S)$ for $n=5$ and the 2-layer approximation method}
\label{fig:ap5}
\end{figure}

\begin{figure}[p]
\begin{center}
\includegraphics[width=10cm]{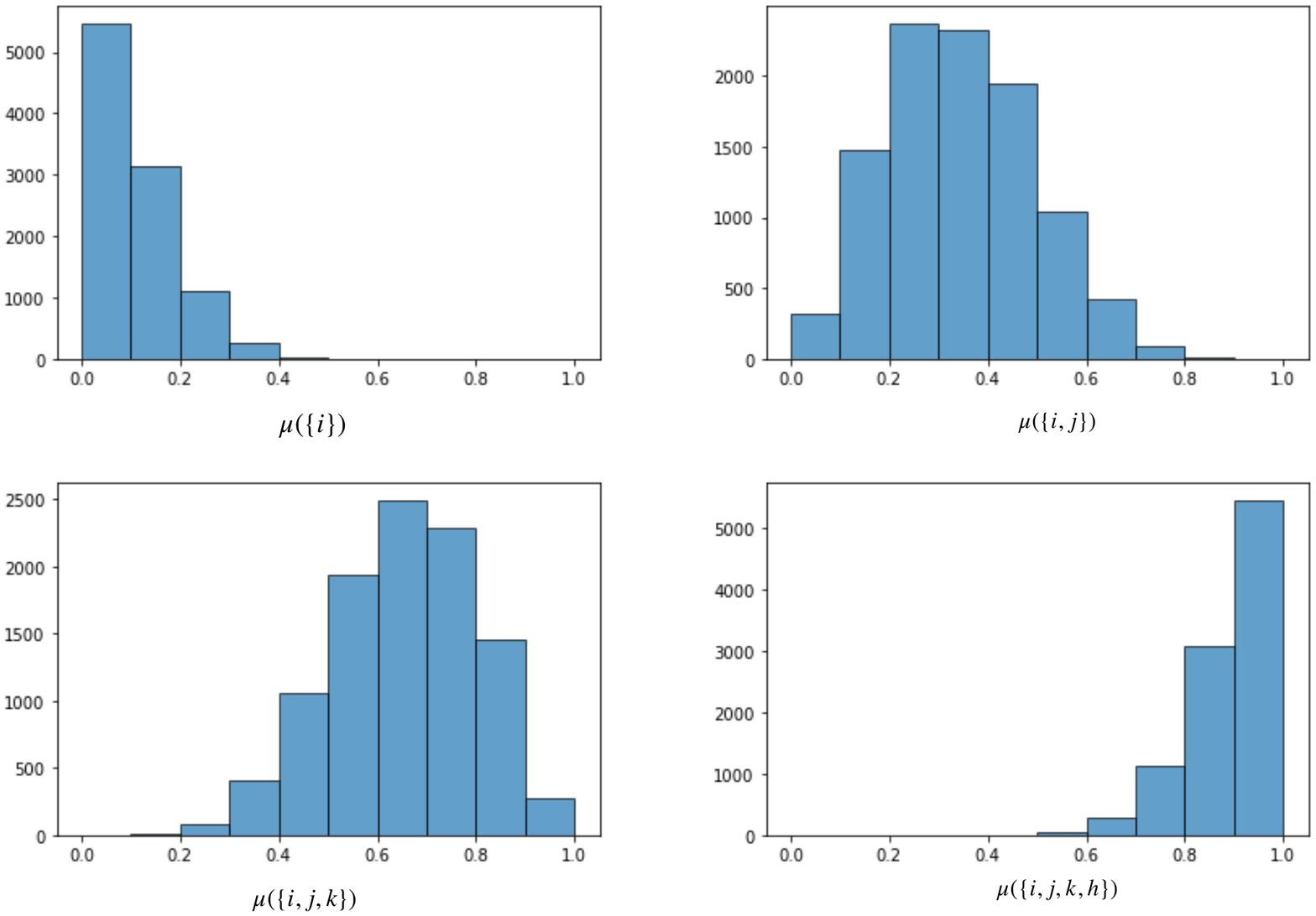}
\end{center}
\caption{Histograms of $\bmu(S)$ for $n=5$ and the Markov chain method}
\label{fig:mc5}
\end{figure}

When $n\geqslant 5$, no comparison with the exact method can be done any more,
and one can only check that the properties of Lemma~\ref{lem:dist} are
satisfied. Figures~\ref{fig:ap5} and \ref{fig:mc5} show that histograms obtained
when $n=5$ for the 2-layer approximation method and the Markov chain method. One
can see that both method perform well, with a slight advantage for the 2-layer
approximation method.

In summary, the 2-layer approximation method and the Markov chain methods haven
similar performance, with a slight advantage for the former. 
Concerning the computation time, however, there is a clear advantage for the
2-layer approximation method (see Table~\ref{tab:perf}). Computations are done
on a 3.2 GHz PC with 16 GB of RAM.
\begin{table}[htb]
\begin{center}
\begin{tabular}{|c|c|c|}
\hline
method & $n=4$ & $n=5$\\
\hline
Two-layer approximation method & 2s & 13s\\
Markov Chain method & 26s& 265s \\ \hline
\end{tabular}
\end{center}
\caption{Comparison of CPU time for generating 10000 capacities}
\label{tab:perf}
\end{table}

\subsection{The centroid}
In order to check if the set of generated capacities is well balanced in the
whole polyhedron $\cC(N)$, we propose to compute its average capacity and to
check if it coincides with the centroid of $\cC(N)$. We recall that for a
convex polyhedron $P$, the centroid can be obtained from a triangulation of the
polyhedron into simplices. The centroid of a simplex is the barycenter of its
vertices, and the centroid of $P$ is the average of the centroids of all
simplices. Note that it differs from the barycenter of the vertices of $P$ in general.

From Section~\ref{sec:rgline}, we know that $\cC(N)$ can be triangulated by the
regions $R_\sigma$, each region corresponding to a linear
extension, and its vertices are given by Eq. (\ref{eq:1a}).

We illustrate the method of computation with $n=3$. Then, $\cC(N)$ is a
6-dimensional polytope, triangulated into 48 simplices corresponding to the 48
linear extensions (see Table~\ref{tab:1}):
\begin{align*}
\sigma_1 &= \{1\},\{2\},\{3\},\{1,2\},\{1,3\},\{2,3\}\\
\sigma_2 &= \{1\},\{2\},\{3\},\{1,2\},\{2,3\},\{1,3\}\\
\sigma_3 &= \{1\},\{2\},\{3\},\{1,3\},\{1,2\},\{2,3\}\\
\sigma_4 &= \{1\},\{2\},\{3\},\{1,3\},\{2,3\},\{1,2\}\\
\sigma_5 &= \{1\},\{2\},\{3\},\{2,3\},\{1,2\},\{1,3\}\\
 \vdots & \quad \vdots
\end{align*}
According to (\ref{eq:1a}), the vertices of $R_{\sigma_1}$ are:
\[
(0,0,0,0,0,0),\ (0,0,0,0,0,1),\ (0,0,0,0,1,1),\ (0,0,0,1,1,1),\ (0,0,1,1,1,1),\
(0,1,1,1,1,1),\ (1,1,1,1,1,1)
\]
where the coordinates of a capacity $\mu$ in $\cC(N)$ are given in the following
order: $(\mu(\{1\})$,$\mu(\{2\})$,$\mu(\{3\})$,$\mu(\{1,2\})$,
$\mu(\{1,3\})$,$\mu(\{2,3\}))$. Hence, the centroid $c_1$ of this simplex is the
barycenter of the 7 above vertices, which yields
\[
c_1 =  \Big(\frac{1}{7}, \frac{2}{7},
\frac{3}{7},\frac{4}{7},\frac{5}{7},\frac{6}{7}\Big).
\]
Similarly, we obtain for $R_{\sigma_2}$ the centroid $c_2$:
\[
c_2  =\Big(\frac{1}{7}, \frac{2}{7}, \frac{3}{7},\frac{4}{7},\frac{6}{7},\frac{5}{7}\Big).
\]
After computing the centroids of $R_{\sigma_1},\ldots,R_{\sigma_{48}}$, the
centroid $c$ of $\cC(N)$ is the average of all these centroids:
\[
c=(0.298, 0.298, 0.298, 0.702, 0.702, 0.702).
\]

Exact computation can be done also for $n=4$, but not for greater values, as the
number of linear extensions is too huge. Note that the centroid, whose
computation is based on linear extensions, inherits the same properties as the
random variables $\bmu(S)$, i.e., denoting by $c(S)$ the coordinate of $c$
pertaining to subset $S$, $c(S)$ depends only on the cardinality of $S$, and
$c(N\setminus S)=1-c(S)$. This was already remarked by Miranda and Combarro
\cite{mico07}. 

\medskip

Table~\ref{tab:resul3} gives experimental results for $n=3$ obtained with 10000
capacities generated by the 2-layer approximation method and the Markov chain
method, then averaging all of them as an estimation of the centroid. The
squared error indicates the total squared error between the exact method and the
considered method.
\begin{table}[htb]
  \begin{center}
    \begin{tabular}{|c|c|c|}\hline
      Method & centroid & squared error\\ \hline
      exact method & (0.298, 0.298, 0.298, 0.702, 0.702, 0.702) & $-$\\
      2-layer approximation &  (0.298, 0.298, 0.296, 0.703, 0.703, 0.702) &
      $6\times 10^{-6}$\\
      Markov chain &  (0.297, 0.297, 0.297, 0.702, 0.703, 0.701) & $5\times 10^{-6}$\\ \hline
    \end{tabular}
  \end{center}
  \caption{The centroid and its estimated values for $n=3$}
  \label{tab:resul3}
\end{table}

Table \ref{tab:resul4} is for $n=4$, and we have generated 10 times 10000
capacities in order to compute the standard deviation for the 2-layer
approximation method and the Markov chain method over the 10
realizations.
\begin{table}[htb]
  \begin{center}
    \begin{tabular}{|p{4cm}|p{8cm}|c|}\hline
      Method & centroid & squared error\\ \hline
      exact method & (0.1810, 0.1810, 0.1810, 0.1810, 0.5000, 0.5000, 0.5000, 0.5000,
      0.5000,  0.5000, 0.8190, 0.8190, 0.8190, 0.8190) & $-$ \\ \hline
      2-layer approximation (mean value)& (0.1821, 0.1823, 0.1822, 0.1825, 0.5010,
      0.5008, 0.5015, 0.5015, 0.5014, 0.5015, 0.8196, 0.8201  0.8204, 0.8200) &
      $2147\times 10^{-8}$\\
      standard deviation & (0.0015,  0.0013,  0.0010,  0.0013,  0.0017,  0.0014,
            0.0018,  0.0011,  0.0017,  0.0009,  0.0015, 0.0012, 0.0011, 0.0008) &\\
      \hline
      Markov chain (mean value)& (0.1812, 0.1804, 0.1808, 0.1813, 0.5002,
      0.5008, 0.5003, 0.5005, 0.5004, 0.5006, 0.8199, 0.8200,  0.8201, 0.8204) &
      $705\times 10^{-8}$\\
      standard deviation & (0.0008, 0.0008, 0.0012, 0.0005, 0.0015, 0.0010,
      0.0017, 0.0013, 0.0016, 0.0011, 0.0010, 0.0014, 0.0009, 0.0009)& \\
      \hline
    \end{tabular}
  \end{center}
  \caption{The centroid and its estimated values for $n=4$}
  \label{tab:resul4}
\end{table}
Both methods show very good performance, with a slight advantage for the Markov
chain method in terms of squared error when $n=4$.

\subsection{Related literature}
Several methods are used in the literature to measure the performance of a
generator. Combarro et al. \cite{codimi13} use the centroid and its properties
of symmetry mentioned above. They compute via an index they define if the
coordinates of the average of the generated capacities are indeed depending only
on the cardinality and if the symmetry $S$ vs. $N\setminus S$ is verified. This,
however, is not sufficient to ensure that the averaga capacity is close to the
centroid.

In \cite{cohudi19}, Combarro et al. propose to measure the total variation distance to the uniform
distribution, that is to compute $\frac{1}{2}\sum_l|p_l - \frac{1}{e(p)}|$ with
$p_l$ the probability of obtaining linear extension $l$ with a given capacity
generator, and $e(p)$ is the total number of linear extensions.

Another method used in \cite{hapi17} is to draw the histogram and calculate the
entropy of the weight $w_{\sigma(i)}$ on each source $x_{\sigma(i)}$ with
$w_{\sigma(i)} = \mu(A_i) - \mu(A_{i-1})$ and $A_i =
\{x_{\sigma(1)},\ldots,x_{\sigma(i)}\}$. However, for the distribution of
  $w_{\sigma(i)}$, we only know that it holds symmetry, but nothing more is
  known about its properties.

\bibliographystyle{plain}
\bibliography{../BIB/fuzzy,../BIB/grabisch,../BIB/general}

\appendix

\section{Study of posets $T_H$ closed under intersection}
We give here a characterization of posets $T_H$ closed under intersection, as
well as the proof of Proposition~\ref{prop:reg}.

We characterize now the form of  any $T_H$ closed
under intersection. To this end, we need to distinguish two cases:
\begin{enumerate}
\item Situation 1: for any node $y$ of the lower layer, $|\prec(y)|\leq 2$.
\item Situation 2: there exists a unique $y$ of the lower layer
  s.t. $|\prec(y)|=h$, with $h$ the number of nodes in the upper layer, and $h\geqslant 3$.
\end{enumerate}
We claim that under the assumption of closure under intersection, there is no
other alternative.
\begin{lemma}\label{lem:2}
For any $T_H$ closed under intersection there is no
other possibility than Situations 1 and 2.
\end{lemma}
\begin{proof}
Let $h$ be the number of nodes in the upper layer.
  We begin by remarking that if a node of the lower layer has $h$ predecessors, it
must be unique as it is the intersection of all nodes of the upper layer.
  
The cases where $h=1,2$ are trivial. For $h=3$, either the number of
predecessors of a lower node is at most 2, or it is 3, so either Situation 1 or
2 arises.

Assume then $h>3$. Suppose there exist a node $T$ at the lower level with
$\prec(T)$ containing at least three nodes, say $T\cup \{1\}, T\cup \{2\}, T\cup
\{3\}$ (therefore $T\subseteq N\setminus \{1,2,3\}$). Assume by contradiction
that there exist $S$ in the upper level such that $T\not\in\succ(S)$. Since
$T_H$ is closed under intersection, there must exist a successor common
to $T\cup \{1\}$ and $S$. As $T\not\subseteq S$, this successor has the form
$(T\setminus \{i\})\cup \{1\}$ for some $i\in T$. Therefore, $S\supseteq
(T\setminus \{i\})\cup \{1\}$. Similarly, $T\cup \{2\}$ and $S$ must have a
common successor, which implies that $2\in S$. Similarly, we have also that
$3\in S$. Then
\[
|S|\geq 3+ |T\setminus \{i\}|=|T|+2,
\]
which is impossible by definition of $T_H$.
\end{proof}

\begin{proposition}
Assume $T_H$ is closed under intersection, let $N':=\bigcup_{x\in \text{upper layer}}
x$, with $|N'|=n'\leqslant n$, and denote by $h$ the number of nodes in the
upper layer, of cardinality $\ell$.

The two situations are characterized as follows:
\begin{itemize}
\item Situation 1: 
  \begin{itemize}
  \item Case 1: $h=1$, i.e., there is only one element in the upper layer;
  \item Case 2: $h\geqslant 2$. Then $\ell=n'-1$, i.e.,  the nodes of the upper
    layer are of the form $N'\setminus \{i\}$. We have $\frac{h(h-1)}{2}$
    nodes in the lower layer having exactly $2$ predecessors.
  \end{itemize}
\item Situation 2: $h=n'-\ell+1$ and $\ell<n'-1$.  Nodes of the upper level are
  of the form $S\cup\{i\}$, and the lower layer contains the node $S$ with $h$
  predecessors, and possibly other nodes with $0$ or $1$ predecessor.
\end{itemize}
\label{prop1}
\end{proposition}

\begin{proof}
\underline{For Situation 1:} 
Suppose we have $h$ nodes in the upper layer, $h\geqslant 2$. As $T_H$ is closed
under intersection, any two nodes in the upper layer have a common successor in the lower layer.
Now, in Situation 1, any node in the lower layer has at most $2$ predecessors.
Hence the $\frac{h(h-1)}{2}$ pairs of nodes $x,x'$ in the upper layer yield
$\frac{h(h-1)}{2}$ distinct elements in the lower layer. Indeed, if they were not distinct, some node in the lower level would have at least $3$ predecessors.

It remains to prove that $\ell=n'-1$. To this aim, let us take two nodes in the
upper layer with common successor the set $S\subseteq N'$. Then these two nodes
must be of the form $S\cup \{i\}$ and $S\cup \{j\}$. If $h=2$, we have
$N'=S\cup\{i,j\}$, which proves the
assertion. Suppose now that $h\geqslant 3$ and consider in the upper layer a
set $T\neq S\cup\{i\},S\cup\{j\}$. We claim that $N' =S\cup\{i,j\}$,  i.e.,
$T\subseteq S\cup\{i,j\}$, which
proves our assertion. Suppose by contradiction that $T$ contains an element $l$
such that $l\not\in S\cup\{i,j\}$. Let us write $T=S'\cup\{l\}$ with $S'\not\ni
l$. As $T_H$ is closed under intersection, $T$ and $S\cup\{i\}$ have a common
successor, which must be $S'$, because $l$ does not belong to the successor and
the successor must have $\ell-1$ elements. Similarly, the common successor of
$T$ and $S\cup\{j\}$ must be $S'$. But then $S'$ has three predecessors, a
contradiction.

\bigskip

\underline{For Situation 2:}
Denoting by $S_1,\ldots,S_h$ the nodes in the upper layer, we know by definition
of Situation 2 that there exists $S$ in the lower layer
s.t. $S=S_1\cap\cdots\cap S_h$. Hence, the nodes in the upper layer have
the form  $S\cup \{i\}$ with $i\in N'\setminus S$. 
Consequently, $h=n'-|S| = n'-\ell+1$, i.e., $\ell=n'-h+1<n'-1$ since $h\geqslant
3$. 

Moreover, any two nodes $S\cup\{i\}$ and $S\cup\{j\}$ of the upper layer have a
unique common successor, which is $S$. Hence, any $T \not= S$ in the lower layer
must have $0$ or $1$ predecessor.
\end{proof}

Of course, symmetric results can be established for the two bottom layers
$B_H$. 

\begin{figure}[htb]
\begin{center}
\begin{tikzpicture}[scale=0.80,node distance=5mm,
loose/.style={
rectangle,minimum size=6mm,rounded corners=3mm,
very thick,draw=red!50, top color=red!10,bottom color=red!10,
font=\scriptsize
},
antichain/.style={
rectangle,minimum size=6mm,rounded corners=3mm,
very thick,draw=green!100, top color=green!30,bottom color=green!30,
font=\scriptsize
},
win/.style={
rectangle,minimum size=6mm,rounded corners=3mm,
very thick,draw=blue!50,
top color=blue!10,bottom color=blue!10,
font=\scriptsize
},
topo/.style={
rectangle,minimum size=6mm,rounded corners=3mm,
very thick,draw=black!50,
top color=black!10,bottom color=black!10,
font=\scriptsize
}]

\tikzstyle{s}=[circle,draw, line width=1pt]

\tikzstyle{s-out}=[circle,draw, double, line width=1pt]

\node (u1) at (0,3) [topo] {$x$};
\node (u2) at (1.5,3) [topo] {};
\node (u3) at (3,3) [topo] {};
\node at (4.2,3) {$\cdots$};
\node (u4) at (5.5,3) [topo] {};
\node (u5) at (6.5,3) [topo] {};
\node (l1) at (-3.75,0) [win] {};
\node at (-2.85,0) {$\cdots$};
\node (l2) at (-2,0) [win] {};
\node (l3) at (-0.5,0) [antichain] {};
\node at (0.35,0) {$\cdots$};
\node (l4) at (1.2,0) [antichain] {};
\node (l5) at (2.2,0) [antichain] {};
\node at (3.2,0) {$\cdots$};
\node (l6) at (4,0) [antichain] {};
\node (l7) at (5.5,0) [loose] {};
\node at (6.4,0) {$\cdots$};
\node (l8) at (7.2,0) [loose] {};
\node (l9) at (8.2,0) [loose] {};
\node at (9.2,0) {$\cdots$};
\node (l10) at (10,0) [loose] {};
\draw [decorate,  decoration = {brace}] (-0.5,3.5) --  (7,3.5);
\node at (3.25,4) {$h$};
\draw [decorate,  decoration = {brace,mirror}] (-4.2,-0.5) --  (-1.5,-0.5);
\node at (-3,-1) {$|I|$};
\draw [decorate,  decoration = {brace,mirror}] (-1,-0.5) --  (1.6,-0.5);
\node at (0.5,-1) {$n_x(1)$};
\draw [decorate,  decoration = {brace,mirror}] (5.1,-0.5) --  (7.6,-0.5);
\node at (6.4,-1) {$n_x(2)=h-1$};
\draw [decorate,  decoration = {brace,mirror}] (5.1,-1.5) --  (10.6,-1.5);
\node at (7.7,-2) {$\frac{h(h-1)}{2}$};
\node at (-2.9,-2.8) {$|\prec(y)|=0$};
\draw [dashed] (-1.25,-0.5) -- (-1.25,-3);
\node at (2.3,-2.8) {$|\prec(y)|=1$};
\draw [dashed] (4.75,-0.5) -- (4.75,-3);
\node at (7.5,-2.8) {$|\prec(y)|=2$};
\draw [->,line width=1pt] (l3) -- (u1);
\draw [->,line width=1pt] (l4) -- (u1);
\draw [->,line width=1pt] (l5) -- (u2);
\draw [->,line width=1pt] (l6) -- (u5);
\draw [->,line width=1pt] (l7) -- (u1);
\draw [->,line width=1pt] (l7) -- (u2);
\draw [->,line width=1pt] (l8) -- (u1);
\draw [->,line width=1pt] (l8) -- (u5);
\draw [->,line width=1pt] (l9) -- (u2);
\draw [->,line width=1pt] (l9) -- (u3);
\draw [->,line width=1pt] (l10) -- (u4);
\draw [->,line width=1pt] (l10) -- (u5);
\end{tikzpicture}
\end{center}
\caption{Situation 1.}
\label{fig:c1}
\end{figure}
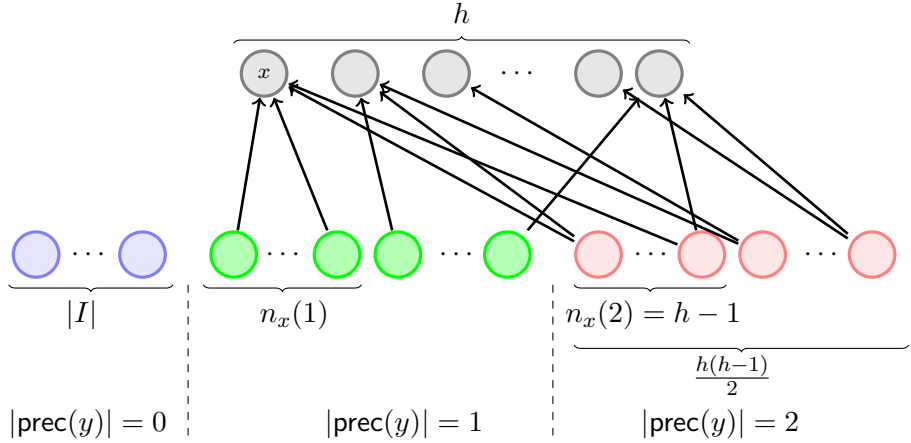

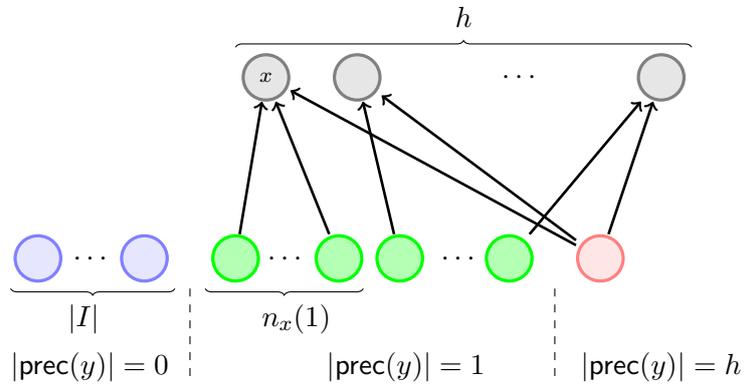
\begin{figure}[htb]
\begin{center}
\begin{tikzpicture}[scale=0.80,node distance=5mm,
loose/.style={
rectangle,minimum size=6mm,rounded corners=3mm,
very thick,draw=red!50, top color=red!10,bottom color=red!10,
font=\scriptsize
},
antichain/.style={
rectangle,minimum size=6mm,rounded corners=3mm,
very thick,draw=green!100, top color=green!30,bottom color=green!30,
font=\scriptsize
},
win/.style={
rectangle,minimum size=6mm,rounded corners=3mm,
very thick,draw=blue!50,
top color=blue!10,bottom color=blue!10,
font=\scriptsize
},
topo/.style={
rectangle,minimum size=6mm,rounded corners=3mm,
very thick,draw=black!50,
top color=black!10,bottom color=black!10,
font=\scriptsize
}]

\tikzstyle{s}=[circle,draw, line width=1pt]

\tikzstyle{s-out}=[circle,draw, double, line width=1pt]

\node (u1) at (0,3) [topo] {$x$};
\node (u2) at (1.5,3) [topo] {};
\node at (4.2,3) {$\cdots$};
\node (u3) at (6.5,3) [topo] {};
\node (l1) at (-3.75,0) [win] {};
\node at (-2.85,0) {$\cdots$};
\node (l2) at (-2,0) [win] {};
\node (l3) at (-0.5,0) [antichain] {};
\node at (0.35,0) {$\cdots$};
\node (l4) at (1.2,0) [antichain] {};
\node (l5) at (2.2,0) [antichain] {};
\node at (3.2,0) {$\cdots$};
\node (l6) at (4,0) [antichain] {};
\node (l7) at (5.5,0) [loose] {};
\draw [decorate,  decoration = {brace}] (-0.5,3.5) --  (7,3.5);
\node at (3.25,4) {$h$};
\draw [decorate,  decoration = {brace,mirror}] (-4.2,-0.5) --  (-1.5,-0.5);
\node at (-3,-1) {$|I|$};
\draw [decorate,  decoration = {brace,mirror}] (-1,-0.5) --  (1.6,-0.5);
\node at (0.5,-1) {$n_x(1)$};
\node at (-2.9,-1.8) {$|\prec(y)|=0$};
\draw [dashed] (-1.25,-0.5) -- (-1.25,-2);
\node at (2.3,-1.8) {$|\prec(y)|=1$};
\draw [dashed] (4.75,-0.5) -- (4.75,-2);
\node at (6.5,-1.8) {$|\prec(y)|=h$};
\draw [->,line width=1pt] (l3) -- (u1);
\draw [->,line width=1pt] (l4) -- (u1);
\draw [->,line width=1pt] (l5) -- (u2);
\draw [->,line width=1pt] (l6) -- (u3);
\draw [->,line width=1pt] (l7) -- (u1);
\draw [->,line width=1pt] (l7) -- (u2);
\draw [->,line width=1pt] (l7) -- (u3);
\end{tikzpicture}
\end{center}
\caption{Situation 2.}
\label{fig:c2}
\end{figure}

Figures~\ref{fig:c1} and
\ref{fig:c2} illustrate Situations 1 and 2, as well as the function $n_x$.

\paragraph{Proof of Proposition~\ref{prop:reg}.}

  \begin{enumerate}
  \item Observe that $n_x(0)=k-\succ(x)$, where $k$ is
    the number of nodes in the lower layer. As $n_x$ is invariant, $\succ(x)$
    does not depend on $x$.
  \item We distinguish two cases: Situation 1 and Situation 2.

1. Suppose we are in Situation 1, i.e., $|\prec(y)|=0$, 1 or 2 for every $y$ in
the lower layer. We have by definition $n_x(0)=k-|\succ(x)|$. As $|\succ(x)|$ is
constant with $x$ by blancedness, $n_x(0)$ does not depend on $x$. Now, as
$T_H[h,k,0]$ is closed under intersection, a given node $x$ has $h-1$ distinct
successors $y$ given by $\succ(x)\cap\succ(x')$ for each $x'\neq x$ in the upper
layer. Therefore $|\prec(y)|=2$ for each of them, so that $n_x(2)=h-1$. Finally,
$n_x(1)=k-n_x(2)-n_x(0)$, which establishes the result.

2. Suppose we are in Situation 2. From Proposition~\ref{prop1}, we know that
$|\prec(y)|=0$, 1 or $h$.  It suffices to prove that $n_x$ is invariant
w.r.t. $x$. We know from Proposition~\ref{prop1} again that there is a unique
node $y$ with $|\prec(y)|=h$, so that clearly $n_x(h)=1$ for every $x$. Similarly
as above, $n_x(0)=k-|\succ(x)|$, which is invariant w.r.t. $x$. Finally,
$n_x(1)=k-n_x(h)-n_x(0)$, which establishes the result.
\item Take $T_H = \{123, 124, 23, 24\}$. Then $n_x$ is invariant but $T_H$ is not
  closed under intersection.
\item Take $H=2^N\setminus\{N,\varnothing\}$, with $|N|=n$. Then $f_x(y)=2$ for
  every $y$ successor of $x$, so that clearly $n_x$ is invariant. 
\end{enumerate}

\end{document}